\documentclass{emulateapj}
\usepackage{graphicx}

\newcommand{\ha}{\mbox{H$\alpha$}}
\newcommand{\oii}{\hbox{[O {\sc ii}]}}
\newcommand{\oiii}{\hbox{[O {\sc iii}]}}
\newcommand{\kms}{$\rm {km}~\rm s^{-1}$}
\newcommand{\hstv}{\hbox{$V_{606}$}}
\newcommand{\hsti}{\hbox{$I_{814}$}}
\newcommand{\etal}{et al.\,}
\newcommand{\solar}{\odot}
\newcommand{\x}{\enspace}

\long\def\comment#1{\relax}

\shortauthors{Weiner et al.}
\shorttitle{DEEP Galaxy Redshift Survey. III.}

\slugcomment{ApJ in press}

\begin{document}

%\twocolumn[

\title{The DEEP Groth Strip Galaxy Redshift Survey. III. \\ Redshift Catalog
and Properties of Galaxies}

\author{Benjamin J. Weiner\altaffilmark{1,2}, 
Andrew C. Phillips\altaffilmark{1},
S.M. Faber\altaffilmark{1}, 
Christopher N.A. Willmer\altaffilmark{1,3}, 
Nicole P. Vogt\altaffilmark{4}, 
Luc Simard\altaffilmark{5},
Karl Gebhardt\altaffilmark{6},
Myungshin Im\altaffilmark{7},
D.C. Koo\altaffilmark{1}, 
Vicki L. Sarajedini\altaffilmark{8},
Katherine L. Wu\altaffilmark{8},
Duncan A. Forbes\altaffilmark{9},
Caryl Gronwall\altaffilmark{10},
Edward J. Groth\altaffilmark{11},
G.D. Illingworth\altaffilmark{1},
R.G. Kron\altaffilmark{12},
Jason Rhodes\altaffilmark{13},
A.S. Szalay\altaffilmark{14},
M. Takamiya\altaffilmark{15}
}

%\email{bjw@ucolick.org}

\altaffiltext{1}{UCO/Lick Observatory, University of California, Santa Cruz, 
Santa Cruz, CA 95064,
{\tt bjw@ucolick.org}}
%{\tt bjw, phillips, faber, cnaw, koo@ucolick.org}}
\altaffiltext{2}{Present address: Department of Astronomy, University
of Maryland, College Park, MD 20742}
\altaffiltext{3}{On leave from Observatorio National, Brasil}
\altaffiltext{4}{Department of Astronomy, New Mexico State University, 
P.O. Box 30001, Las Cruces, NM 88003
%, {\tt nicole@nmsu.edu}
}
\altaffiltext{5}{
%Canadian Astronomy Data Centre, 
Herzberg Institute of Astrophysics, National Research Council of Canada,
5071 W. Saanich Rd., Victoria, BC V9E 2E7, Canada
%, {\tt luc.simard@nrc.ca}
}
\altaffiltext{6}{Department of Astronomy, University of Texas, 
Austin, TX 78723
%, {\tt gebhardt@astro.as.utexas.edu}
}
\altaffiltext{7}{Astronomy Program, SEES, Seoul National University,
Seoul, Korea
%, {\tt mim@astroim.snu.ac.kr}
}
\altaffiltext{8}{Department of Astronomy, University of Florida, 
Gainesville, FL 32611
%, {\tt vicki, klwu@astro.ufl.edu}
}
\altaffiltext{9}{Centre for Astrophysics and Supercomputing, 
Swinburne University, Hawthorn, VIC 3122, Australia}
%, {\tt dforbes@swin.edu.au}}
\altaffiltext{10}{Department of Astronomy and Astrophysics, Pennsylvania 
State University, University Park, PA 16802}
%, {\tt caryl@astro.psu.edu}}
\altaffiltext{11}{Department of Physics, Princeton University,
Princeton, NJ 08544}
%, {\tt groth@pupgg.princeton.edu}}
\altaffiltext{12}{Yerkes Observatory, 373 W. Geneva St.,
Williams Bay, WI 53191}
%, {\tt rich@oddjob.uchicago.edu}}
\altaffiltext{13}{Jet Propulsion Laboratory, California Institute of 
Technology, 4800 Oak Grove Drive, Pasadena, CA 91109}
%, {\tt rhodes@astro.caltech.edu}}
\altaffiltext{14}{Department of Physics and Astronomy, Johns Hopkins 
University, Baltimore, MD 21218}
%, {\tt szalay@jhu.edu}}
\altaffiltext{15}{Institute for Astronomy, 640 North A'ohoku Place \#209,
Hilo, Hawaii 96720}
%, {\tt takamiya@hawaii.edu}}

\begin{abstract}

The Deep Extragalactic Evolutionary Probe (DEEP) is a series of
spectroscopic surveys of faint galaxies, targeted at understanding the
properties and clustering of galaxies at redshifts $z \sim 1$.  We
present the redshift catalog of the DEEP 1 Groth Strip pilot phase of
this project, a Keck/LRIS survey of faint galaxies in the Groth Survey
Strip imaged with HST WFPC2.  The redshift catalog and data, including
reduced spectra, are made publicly available through a Web-accessible
database.  The catalog contains 658 secure galaxy redshifts with a
median $z=0.65$.  The distribution of these galaxies shows large-scale
structure walls to $z \sim 1$.  
We find a bimodal distribution in the galaxy color-magnitude diagram
which persists to the same distance.  A similar color division has
been seen locally by the SDSS survey 
and to $z \sim 1$ by the COMBO-17 survey.
The HST imaging allows us to measure structural properties of the galaxies, 
and we find that the color division corresponds generally to a structural
division.  Most red galaxies,  $\sim 75\%$,
are centrally concentrated, with a red
bulge or spheroidal stellar component, while blue galaxies usually
have exponential profiles.  However, there are two subclasses of red
galaxies that are not bulge-dominated: edge-on disks and a second
category which we term diffuse red galaxies (DIFRGs).
Comparison to a local sample drawn from the RC3 suggests that distant
edge-on disks are similar in appearance and frequency to those at low
redshift, but analogs of DIFRGs are rare among local red galaxies.  
DIFRGs have
significant emission lines, indicating that they are reddened mainly by
dust rather than age.  The DIFRGs in our sample are all at $z>0.64$,
suggesting that DIFRGs are more prevalent at high redshifts; they may
be related to the dusty or irregular extremely red objects beyond
$z>1.2$ that have been found in deep $K$-selected surveys.  
We measure the color evolution of both red and blue galaxies by
comparing our $U-B$ colors to those from the RC3.  For red galaxies,
we find a reddening of only 0.11 mag from $z\sim 0.8$ to now, about half
the color evolution measured by COMBO-17.
Larger, more carefully defined
samples with better colors are needed to improve this measurement.
Reconciling evolution in color, luminosity, mass, morphology, and
star-formation rates will be an active topic of future research.

%(Blanton \etal 2003)
%(Bell \etal\ 2003a).  
% Bell \etal\ (2003a).  

\end{abstract}


% ]

\section{Introduction}

Redshift surveys of the distant universe enable the study
of the properties of galaxies at an earlier age, {\it in situ}.
Large nearby galaxy surveys such as the 2dF Galaxy Redshift
Survey (Colless \etal 2001) and Sloan Digital Sky Survey 
(Abajazian \etal\ 2003) are now completing the collection of
huge samples to determine the structural and stellar-population 
parameters of local galaxies to high accuracy.  At the same time,
the development of more powerful multi-object spectrographs on
large telescopes is allowing high-redshift surveys to collect 
truly large samples of distant galaxies, with numbers and
detail useful for comparison to the nearby surveys.

The DEEP Extragalactic Evolutionary Probe (DEEP) is
a series of large-scale galaxy redshift surveys aimed at
studying the evolution of galaxies and large-scale
structure to redshifts $z \sim 1$,
roughly half the present age of the Universe.
The DEEP program is in two parts.  The second part, DEEP2, is a large
survey of $\sim$65,000 galaxies that is now ongoing with
the DEIMOS spectrograph on the Keck 2 telescope.  
The present paper reports results from DEEP1, an earlier,
smaller pilot program using the Low Resolution Imaging Spectrograph
(LRIS) on Keck 1.  The data reported here cover a subset of
the HST Groth Survey Strip (GSS), a mosaic of 28 HST WFPC2
pointings covering 127 arcmin$^2$
on the sky in the \hstv\ and \hsti\
filters (Groth \etal\ 1994; Rhodes, Refregier \& Groth 2000). 
HST photometry of the GSS reaches to a depth of
$\hstv = 26$ and $\hsti = 25$ (5$\sigma$).  The construction
of the spectroscopic sample is discussed in Paper I (Vogt \etal\ 2004
in preparation),
and the measurement of photometric structural parameters is
discussed in Paper II (Simard \etal\ 2002).

We present a catalog containing 658
galaxy redshifts, with median $z = 0.651$,
including 620 galaxies from our spectroscopy and 38 galaxies
from other sources (including 33 from the Canada-France Redshift
Survey, Lilly \etal\ 1995a). 
Galaxies were selected generally to $(V+I)/2 = 24$ mag (Vega).
We use this catalog to discuss the properties of
galaxies and large-scale structure to $z \sim 1$,  
with particular focus on the bimodal distribution of 
properties of the distant galaxy population.

This bimodality has been discovered recently for local galaxies
in Sloan Digital Sky Survey data (e.g., Strateva \etal\ 2001, 
Hogg \etal\ 2003, Kauffmann \etal\ 2003) and for
distant galaxies to $z\sim1$ in the COMBO-17
photometric redshift survey (Bell \etal\ 2003a,
hereafter B03).  Among local galaxies, bimodality is evidenced by a
fairly rapid change in galaxy properties near a stellar
mass of $3 \times 10^{10}$ M$_{\solar}$  (Kauffmann \etal\ 2003).
Above that mass, galaxies are generally red with large spheroids;
below that, they are blue and disk-dominated.

An easy way to visualize the bimodality is
in the color-magnitude diagram, which shows two separate color
sequences, a broad blue sequence and a narrower
red sequence offset to brighter magnitudes and higher masses.
This is shown well, as a function of
redshift, by B03.  We confirm their finding of
a persistent bimodal distribution here and
use the particular features of our data to further investigate
the properties of both red and blue galaxies. 

\subsection{Previous redshift surveys}

Two published redshift surveys that are comparable in size and depth
to DEEP1/GSS are the Canada-France Redshift Survey (CFRS, Lilly
\etal\ 1995a) and the Caltech Faint Galaxy Redshift Survey (CFGRS,
Cohen \etal\ 2000).  The CFRS is a magnitude-limited survey to
$I_{AB}=22.5$ containing 591 galaxy redshifts with median $z=0.56$
(Crampton \etal\ 1995).  CFRS is roughly 1.5 mag shallower than the
nominal limit of DEEP1, but the median redshifts are similar due to 
differences in the detailed sampling of the magnitude range and
to difficulties in obtaining redshifts beyond $z\sim 1.1$ (see
Section \ref{sec-zfail}).  The CFRS is in several fields, one
overlapping the Groth Strip.  The spectral resolution is 40 \AA; 
a subsample of 246
objects have HST imaging, all with $I$ and many also with $B$ or $V$
(Brinchmann \etal\ 1998).

The major region surveyed by the CFGRS is the HST Hubble Deep Field,
where 602 galaxy redshifts are compiled to $R=24$ in the HDF and
$R=23$ in the Flanking Fields (Cohen \etal\ 2000).  The CFGRS combines
redshifts from several authors; most spectra were taken using the
LRIS spectrograph (Oke \etal\ 1995), the same instrument used for
DEEP1.  The typical spectral resolution is 10 \AA, and the median
redshift is about 0.7.  Most galaxies have HST imaging, roughly one
quarter being imaged deeply with multiple filters in the HDF, and
three quarters imaged lightly in $I$ in the Flanking Fields.

A third survey to which we will make repeated reference is
the COMBO-17 photometric redshift survey,
in four fields (Wolf \etal\ 2003; B03).
Unlike traditional spectroscopic surveys, 
this program used intermediate-band photometry
in 17 filters to measure restframe colors and
redshifts accurate to $\Delta z \sim 0.05$ for 25,000 galaxies from
$z = 0.2-1$.  Recent papers have presented
morphological results based on HST ACS imaging
of one of these fields, called GEMS (Rix \etal\ 2004).
%In Willmer \etal\ 2004, we compare DEEP
%luminosity functions to those measured by COMBO-17.  

\subsection{The DEEP GSS Survey}

The DEEP1 Groth Strip data have three distinguishing features
compared to previous efforts.  The spectral
resolution is significantly higher, 
2.9 \AA\ (FWHM) in the blue and 4.2 \AA\ in the red, yielding
a typical velocity resolution of 160 \kms.  This is high enough
for unambiguous identification of the \oii\ 3727 doublet and
to permit kinematic linewidth and rotation 
measurements, which will be presented 
in separate papers (Weiner \etal\ in preparation, Vogt \etal\
in preparation).  The second feature is 
HST imaging in both $V$ and $I$ filters, enabling
measurement of uniform restframe colors and structural parameters
for nearly all objects.  
The third feature is a single, nearly continuous
field with long dimension 40\arcmin\ on the sky,
or 38 comoving Mpc at $z = 1$.  These features, particularly 
the HST imaging, are central to our discussion.

Several papers related to the DEEP1 GSS survey have already 
been published.  Koo \etal\ (1996) presented a small
number of early redshifts.  Vogt \etal\ (1996, 1997)
presented the Tully-Fisher relation for a few resolved
disk galaxies.  Simard \etal\ (1999)
measured the evolution of the size-luminosity relation, and
Simard \etal\ (2002) measured the photometry and structural 
parameters used in this paper.   Of particular relevance
is Im \etal\ (2002), who identified normal E/S0
galaxies to $I<22$ in the present redshift sample and observed
that they populate a narrow color sequence in observed
$V-I$ versus redshift.  This was used to estimate photometric
redshifts for a larger number of E/S0 galaxies
without spectra, from which the luminosity function
of early-type galaxies was computed.
We confirm the narrow range in color of early-type
galaxies here and extend the morphological conclusions of
Im \etal\ (2001) to fainter galaxies.

All HST Johnson-Cousins magnitudes used in this paper are on
the Vega system.  We use a $\Lambda$CDM cosmogony with
$h=0.7, \Omega_M=0.3$, and $\Omega_\Lambda=0.7$.

% new section 2

\section{Observations}

\subsection{Sample}

A series of spectroscopic samples was selected from a photometric catalog
produced from HST \hstv\ and \hsti\ imaging, as described in Paper I (Vogt
\etal\ 2004 in preparation).  The limiting magnitude for most objects 
was $(V+I)/2 = 24$, based on $1\arcsec.5$ diameter aperture magnitudes.
%
% Ed Groth published two BAAS abstracts on the GSS:
% 
% Number Counts and the Angular Correlation Function of an HST Survey
% \bibitem[Rhodes, Groth, \& WFPC1 Id Team(1997)]{1997AAS...191.0301R} 
% Rhodes, J., Groth, E., \& WFPC1 Id Team 1997, Bulletin of the American 
% Astronomical Society, 29, 1207 
% 
% A Survey with the HST
% \bibitem[Groth et al.(1994)]{1994AAS...185.5309G} Groth, E.~J., Kristian, 
% J.~A., Lynds, R., O'Neil, E.~J., Balsano, R., Rhodes, J., \& WFPC-1 IDT 
% 1994, Bulletin of the American Astronomical Society, 26, 1403 
%
The different samples were designed to explore a range of
scientific programs.  In addition to a general magnitude limited sample,
several types of objects were prioritized for selection from the photometric
catalog.  These included: elongated candidates for spatially resolved velocity
profile extraction, morphologically peculiar objects, extremely blue or red
galaxies ($V-I < 0.5$ or $V-I > 1.75$), and objects with photometric redshifts
larger than $z = 2$.
% 
% $V-I < 0.5$ or $V-I > 1.75$ within the inner $1^{\prime\prime}$, but 
% this is covered in paper 1
% 
In addition to the primary targeted objects, we acquired 
and extracted spectra for 
objects which fell serendipitously within the slit,
and stars placed on slitmasks purely for use in aligning the mask on
the sky.

The set of spectroscopic targets is therefore not a strictly random sampling
of the photometric catalog.  In practice, however, the observed objects sample
the range of apparent color and magnitude in the photometric catalog fairly
evenly in the range $(V+I)/2 \le 24$ (Paper I, Vogt \etal\ 2004 in 
preparation).  Combining all objects placed within slits and the objects 
used for alignment purposes, usable spectra were obtained for a total of 
813 different objects.

\subsection{Mask design}

The spectra were taken with the LRIS spectrograph (Oke \etal\ 1995) at the
Keck telescopes in successive spring seasons from 1995 through 1999.
Observations were spread over nine observing runs and 25 nights, in
conjunction with several other spring fields.  In total, 36 slitmasks were
observed in the GSS, where each mask contained typically 30 to 50 slits and
several alignment stars.

The first set of ten masks was populated with targets by using printed images
of the target fields and transparent overlays, while the rest were designed
using interactive, computerized selection; see
Vogt \etal\ (2004, in preparation).  The number of objects per
mask increased with time, as we determined the minimum slit length
to obtain high quality background-subtracted spectra for objects
of a given magnitude and color.  Following target and alignment star
selection, all masks were modeled with the {\tt ucsclris} IRAF package written
by A.C. Phillips to account for specific observational constraints
(e.g., anamorphic corrections) and to create instructions for milling the mask.
Nearly all masks were milled with slits of width $1^{\prime\prime}$, with a
slit length ranging from $8^{\prime\prime}$ to $12^{\prime\prime}$.  After the
first year, a small fraction of all slits was tilted at an angle (between
$0^{\circ}$ and $30^{\circ}$) relative to the position angle of the mask,
in order to trace along the major axis of a spatially elongated object or 
to capture two objects within one slit.

% These masks have typically 30 slits per mask.  Subsequent masks were designed
% using interactive, computerized selection to place objects on the masks, and
% have typically 40-45 slits per mask.

Certain objects, typically those at magnitudes $(V+I)/2 > 23$; those with
extremely red colors; those for which a redshift could not be determined after
initial observations; and spatially extended objects 
requiring higher S/N, were
placed on one, three, or (rarely) more additional slitmasks to obtain
more exposure time.

\subsection{Observations}

Each mask was observed in turn with a blue (typically 900
l~mm$^{-1}$/5500 \AA) and a red (600 l~mm$^{-1}$/7500 \AA) grating,
for a combined spectrum covering the range 5000 \AA\ to 8200 \AA.  We obtained
a baseline of
$2 \times 1500$ seconds exposure per grating per mask, and were able to 
keep the airmass below 1.3.
The spectral resolutions, measured by fitting multiple night sky OH lines in
reduced, combined and extracted spectra, are $\sigma = 1.25$ \AA\ in the blue
spectra and 1.8 \AA\ in the red spectra, or FWHM = 2.9 and 4.2 \AA\
respectively.

A matched flat fielding procedure was developed for the later data.  At the end
of the night, the telescope was slewed to the same position and the
spectrograph rotated to the same position angle as used for the
nighttime data exposure.  The grating tilt was iteratively adjusted to place a
given wavelength at the same pixel as in the nighttime exposure, 
%by comparison of arc lines with night sky OH lines, 
and then the flat field image was taken.
This procedure reduced the effect of CCD fringing and
improved the sky subtraction and data quality in
the red portion of the spectrum, beyond 7000 \AA.

\subsection {Data Reduction}

Two programs, each crafted for 2-D multislit spectral reductions, were
used to reduce these data.  Early masks 
were reduced using Dan Kelson's {\sc Expector} software (Kelson \etal\ 2000).
%supplemented by a few other miscellaneous codes.  
Later masks were reduced using the IRAF {\sc Redux} package written by one 
of the authors [ACP].  A few masks were reduced using
both of these packages, yielding comparable results.

Both sets of software perform the same group of operations:
bias subtraction and trimming of overscan region;
``cosmic ray'' removal; bad pixel replacement; flat fielding;
removal of distortion in the spatial direction;
linearization in wavelength; and correction for a ``slit function'' 
(removal of the effects of non-uniform slit width).

The primary difference between them is the manner in which calibrations
(slitlet location, wavelength) are determined.
{\sc Expector} requires the user to locate slit edges,
%for these data,
%this was done by the user interactively setting the slit edges on a
%spatially-rectified image through the mask.  
but automates wavelength calibration, using a
cross-correlation algorithm.  A wavelength calibration is produced
for each spatial position along
each individual slitlet, using night sky lines and, if necessary,
arc spectra in the blue, taken through each mask.
%These calibrations are then applied on a
%slitlet-by-slitlet basis to the science data.

{\sc Redux} performs its calibrations in a very different way.  The
program uses an optical model of the spectrograph consisting of three
parts: (1) a mapping of mask location to grating input angles; (2) a
transform into and out of the grating coordinate system, where the
grating equation is used to describe the dispersion and
line-curvature; and (3) a mapping of grating output angles to the
detector.  This reduces the effective unknowns to two for each mask:
an error in the grating tilt and an error in the grating
``pitch.'' (A third grating variable, the rotation about the grating
normal, is solved using arc spectra taken through a special mask and
is assumed constant for a given observing run.)  The software uses the
model and the mask design to predict location of slit edges and bright
night sky lines; the offset in the slit locations and the night sky
lines then provide the solution for the two unknown variables.  The
grating tilt error is calculated for each individual
slitlet to provide more precise zero-points in wavelength.  
%With these variables determined, 
The location of each slitlet and the wavelength solution within 
each slitlet are then known to quite good precision for the entire image.

Both {\sc Expector} and {\sc Redux} have advantages and shortcomings.
{\sc Expector} performs empirical calibrations for each mask using
spectra taken through that mask, but when arcs are used
the significant flexure in LRIS means that the calibrations
may not be as accurate for the science data, and offsets
must be added.  In addition,
sparse regions in the arc spectra (typical in the blue) cause
uncertainties in the wavelength solution.
{\sc Redux}, on the other hand, is limited by the precision of 
the optical model.  Using combined data sets with a variety of gratings
and spanning large ranges in grating tilt, typical {\it rms} errors
are $\sim$0.6 px (0.13 arcsec spatially; in dispersion, 0.51 \AA\ and
0.77 \AA\ in the blue and red, respectively), and the errors are
rarely worse than $\sim$1--1.5 px in the very corners of the image.
For most science spectra, the results are better than this, particularly 
as the wavelength scale is zero-pointed using night sky lines.

Fringing in the red has always been a difficult problem due to the
strong night-sky lines in the OH ``forest''. Given the flexure in
LRIS, it was often found that applying a ``fringe frame'' correction
exacerbated the problem rather than reducing it.  Thus, in the earlier
data, fringe frames were only employed if they improved
night-sky subtraction.  This problem was solved in later
data by taking red flat-fields at the end of the night, matched 
in wavelength to the science image as described above.
In these data, fringe frame
corrections were always applied and always produced good results.

Individually-reduced 2-D linearized and rectified spectral images for
each object were collected for all observing runs and combined using a
biweight estimator, resulting in two final combined 2-D images, one 
each for the red and blue gratings.  One-dimensional spectra were then
extracted from these combined images, with a minimum object width of
1.5 arcseconds. 

% end of new section 2

\section{Redshift Catalog}

\subsection{Redshift Determination}

We inspected the 
1-D and 2-D spectra by eye to find spectral features.  For all
candidate redshifts, we verified the existence of features by
visual inspection and assigned a quality code.
For galaxies at $0<z<1.4$, the primary features are nebular emission
lines, principally \oii\ 3726 3729, H$\beta$ 4861, \oiii\ 4959 5007,
and H$\alpha$ 6563; for absorption-line objects the strongest
features are Ca {\sc II} H+K 3933 3969, the CH G-band at 4304 \AA,
Balmer lines, and the 4000 \AA\ break.  Some high-$z$ objects are
identified by interstellar absorption lines of Mg {\sc II} 2796 2803
and Fe {\sc II} 2586 2600.  Cool, late type stars are generally
identified by absorption bands; hotter stars often have featureless
spectra in our wavelength region and are more difficult to identify.

We consider a redshift secure if at least two features are detected.
Since the spectral resolution is high, many galaxies have clearly 
resolved \oii\ doublet emission, which is counted as two features.
Objects with a secure redshift from two or more features are
given a redshift quality A in column 5 of Table \ref{table-catalog}.

Some galaxies have a broad emission line consistent with the 
\oii\ doublet separation of 220 \kms, but not resolved into
two components.  These galaxies have an almost certain redshift,
since the appearance of broad \oii\ in the 2-D spectra is
distinctive, and there is generally no other convincing candidate
for the emission line.  These and other objects with a nearly 
certain redshift but without two strong distinct features are ``almost 
secure'' and are given a quality B in column 5 of Table \ref{table-catalog}.

All redshifts were checked by at least two people (Phillips,
Sarajedini, Koo, Weiner) and disagreements on redshift and 
quality were resolved.  Final redshift qualities were systematized
by one of us (BJW) to assure uniformity.  We estimate that
quality A have $>99\%$ confidence and quality B have $>90\%$ confidence.
Our analysis of the galaxy sample below uses both the quality 
A and B objects.

A few objects have a single feature, very weak features, or
other spectral properties which suggest a redshift, but the
redshift is by no means certain.  
We do not count these as identified or use these objects in the 
analysis of galaxy properties below.  The uncertain redshifts are
not listed in Table \ref{table-catalog} but can be retrieved 
from our Web database.
There are very few objects with a single strong spectral feature that 
yields an ambiguous redshift.
%\comment{(Are there ANY other than broad line QSOs???)}

Objects for which a redshift, 136 in total, could not be determined with 
confidence are coded with quality F in column 5 of 
Table \ref{table-catalog}.  These include objects with low signal-to-noise
and objects with fairly high S/N but no identifiable features.  
A small number
of objects, $\sim 20$, fell off the slit, off the detector, or had
otherwise unusable data; these objects are not listed in Table
\ref{table-catalog} and are not counted in our completeness statistics.

%\begin{table*}[t]
%%\begin{table}[ht]
%\caption{Redshift catalog for the DEEP 1 Groth Strip Survey}
%\label{table-catalog}

%%\psfig{file=fig.v4/tab1.first50.ps,width=5.5truein}
%\includegraphics[width=5.5truein]{fig.v4/tab1.first50.ps}
%\vspace{-0.75in}

%\input table1.v2/tab1.input.dlx.tex

%%%%%%%%%%%%%%%%%%%%
% try to do this as rotated deluxetable

%\begin{deluxetable}{lrrrrrrrrrrrrrrrrr}
\begin{deluxetable}{lrrrrrrrrrrrrrrrrrr}
 
\tablecaption{Redshift catalog for the DEEP 1 Groth Strip Survey
\label{table-catalog}
}
\tablecolumns{19}
\tablewidth{0pt}
\tabletypesize{\scriptsize}

% use this up to the \vspace for emulateapj, since it can't rotate the table
% as \rotate doesn't exist in emulateapj
\label{table-catalog}
\hspace{-0.5in}
\includegraphics[width=5.5truein]{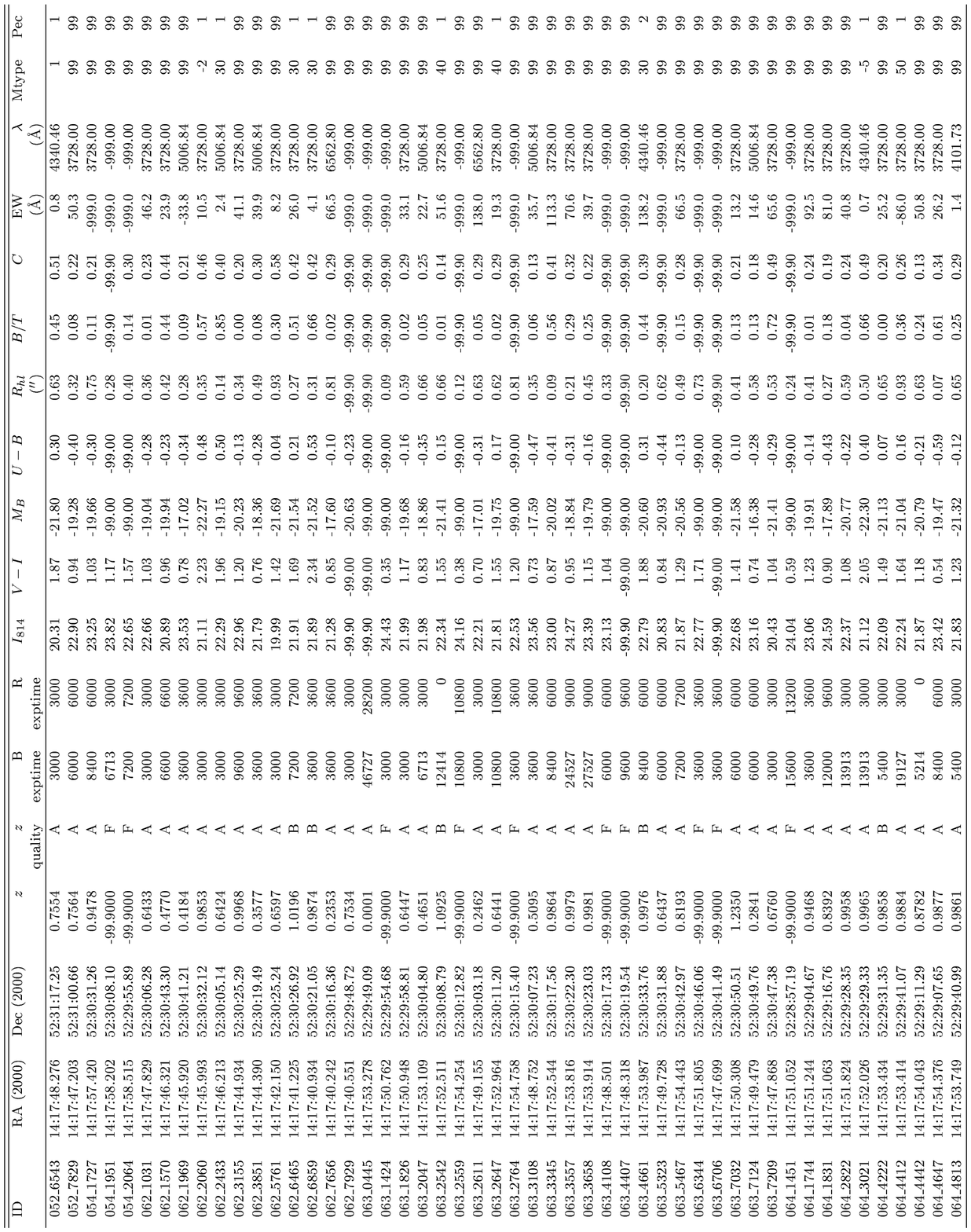}
\vspace{0.5in}

%% use this for AAStex, up to the \enddata
%\setlength{\tabcolsep}{0.04in}
%\rotate
%\tablehead{
%ID  &  RA (2000)  &  Dec (2000)  & $z$ & $z$ & $z$ &  B  & R  & $I_{814}$ & $V-I$ & $M_B$  & $U-B$ & $R_{hl}$ & $B/T$ & $C$  & EW & $\lambda$ & Mtype & Pec \\
%   &   &   &  & quality & source &  exptime & exptime &  &  &  &  & (\arcsec) &  &  & (\AA) & (\AA) &  &  \\
% }
%\startdata
%
%%\input table1.final.first25.data
%\input tab1.tex
%
%\enddata

\tablecomments{The complete version of this table is in the electronic edition of
the Journal.  The printed edition contains only a sample.}

\tablenotetext{1}{Object ID, see Simard \etal\ (2002).}
\tablenotetext{2,3}{RA and declination, Simard \etal\ (2002).}
\tablenotetext{4}{Redshift measured from DEEP spectra or other sources.}
\tablenotetext{5}{Redshift quality. A=secure, B=nearly secure,
F=no redshift; see text.}
\tablenotetext{6}{Redshift source: 1=CFRS (Lilly \etal\ 1995a), 
2=Brinchmann \etal\ (1998), 3=Hopkins \etal\ (2000), 4=DEEP.}
\tablenotetext{7,8}{Exposure times for blue and red spectra in seconds.}
\tablenotetext{9,10}{Total $I$ and $V-I$ (Vega) from Simard \etal\ (2002).}
\tablenotetext{11,12}{Restframe $M_B$ and $U-B$ (Vega), determined using
K-correction procedure described in text.}
\tablenotetext{13}{Half-light radius of the photometric model fit
from Simard \etal\ (2002).}
\tablenotetext{14,15}{Bulge-to-total ratio and concentration parameter
of the photometric model.}
\tablenotetext{16,17}{EW and wavelength of strongest emission line.}
\tablenotetext{18,19}{Morphological type (see Table \ref{table-redmorph})
and peculiarity codes for red galaxies:
1-normal, 2-peculiar, 3-interacting, 99-not classified; see text.}

\end{deluxetable}

%\end{table*}
%\end{table}

% This is to try to get it to flush all the floats (tables and figures)
% Putting \clearpage here works for aastex
%\clearpage

The high resolution of the LRIS spectra allowed us to determine
redshifts to a few parts in $10^{-4}$.  We quantified redshift
accuracy by comparing to observations of the Groth Strip
from the ongoing DEEP 2 survey with the DEIMOS spectrograph,
at even higher spectral resolution.  The DEEP2 redshifts are
known to be accurate to 30 \kms.
From 140 galaxies with both DEEP1 and DEEP2 redshifts, we find
an offset of (DEEP1-DEEP2) $=9$ \kms\ and an RMS $=51$ \kms.
The RMS error of DEEP1 redshifts is thus $\sigma_V = 42$ \kms; in 
redshift it is $\sigma_z = 2.3 \times 10^{-4}$, with a possible 
but small systematic offset.

A portion of the redshift catalog is shown 
in Table \ref{table-catalog}.  The table contains an entry for every
object for which a usable spectrum was actually obtained, including alignment
stars but not including objects that fell off the slit or where the
data were completely unusable.  
47 objects with redshifts from non-DEEP sources (Lilly \etal\ 1995a;
Brinchmann \etal\ 1998; Hopkins \etal\ 2000)are listed 
at the end of the table.  
Only the first page of the table is reproduced in the printed journal.
The full table is available in the electronic version, and the catalog
and data, including extracted spectra, are publicly available through
a World Wide Web interface at \url{http://saci.ucolick.org}.

41 objects with redshifts fall on or near the edges of the area 
imaged with HST, or next to bright objects, and their total
magnitudes and structural parameters could not be measured with {\tt gim2d}.
These do not have $V$ or $I$ magnitudes listed in Table \ref{table-catalog},
although for 31 we were able to compute aperture magnitudes and
infer restframe magnitudes.
Values which are unmeasured, such as for objects without magnitudes
or redshifts, are flagged by clearly out-of-range values, typically 
-99 or -999.

% This is to try to get it to flush all the floats (tables and figures)
% Putting \clearpage here works for emulateapj
\clearpage

The columns in Table \ref{table-catalog} are:

% need to make table match these

Column 1.  Object ID, referring to WFPC2 chip number and pixel position;
see Paper I.

Columns 2-3.  RA and Dec, epoch J2000.

Column 4.  Redshift. 
%``INDEF'' indicates that no redshift could be determined.  
Redshifts noted ``a'' are from the CFRS (Lilly
\etal \ 1995a), ``b'' are from Brinchmann \etal\ (1998),
and ``c'' are from Hopkins \etal\ (2000).

Column 5.  Redshift quality.  ``A'' is a secure redshift based on two
or more spectral features.  ``B''
is an almost-secure redshift based on one feature such as a broad line
that is almost certainly \oii\ 3727.  ``F'' indicates
a failure to obtain a redshift.  ``A'' and ``B'' are secure
enough to be used in the analysis in the remainder of this paper.

Column 6.  Redshift source: 1=CFRS (Lilly \etal\ 1995a), 
2=Brinchmann \etal\ (1998), 3=Hopkins \etal\ (2000), 4=DEEP.

Column 7-8.  Exposure times for red and blue spectra.

Column 9.  Total $I_{814}$ magnitude from structural models fit
to the $I$ image using {\tt gim2d} by Simard \etal\ (2002).

Column 10.  $V_{606} - I_{814}$ color. $I_{814}$ is from Column 9;
$V_{606}$ is the analogous magnitude fitted to the $V$ image. 
For most (602) objects, the ``simultaneous'' fit quantities from Simard
\etal\ (2002) are used, in which $V$ and $I$ fits are constrained
to have the same size parameters.  For 13 objects, these are unavailable,
and the ``independent'' fits are used.  

Column 11.  Absolute magnitude $M_B$ based on apparent magnitude
and color using the K-correction procedure described below.
For 615 objects, {\tt gim2d} total apparent magnitudes are used to
infer $M_B$; for 31 objects, total magnitudes were not available and 
aperture apparent magnitudes were used.

Column 12.  $U-B$ restframe color, derived through the K-correction
procedure described below.

Column 13.  Half-light radius 
$R_{hl}$ in arcsec from the structural model fit to the $I_{814}$
image by Simard \etal\ (2002).  This is the half-light radius
of the {\it face-on} reconstructed model before PSF convolution.
It is more physically
meaningful than the raw half-light radius in circular apertures and is
larger than the latter for edge-on galaxies.

Columns 14 and 15.  Bulge-to-total $B/T$ ratio and central concentration
$C$ from the $I_{814}$ image by Simard \etal\ (2002).

Columns 16 and 17.  Equivalent width and wavelength 
of the strongest emission line.

Column 18 and 19.  Morphological type and peculiarity code
of red galaxies, as defined in Section \ref{sec-morphology}.

Table \ref{table-zstat} summarizes the numbers of spectra taken and
redshifts identified in the DEEP GSS redshift survey.  A small number
of additional redshifts are provided by other observations in the 
Groth Strip, mostly by the CFRS (Lilly \etal\ 1995a), 
and these are listed at the end of
Table \ref{table-catalog} and counted in
Table \ref{table-zstat}.  They are excluded
from the following discussion of completeness.  

\begin{table}[t]
\begin{center}
\caption{Redshift statistics for the Groth Survey Strip}
\label{table-zstat}
\begin{tabular}{lrrrr}
\tableline\tableline
          &           & \multicolumn{3}{c}{Quality} \\
          &     total &    A  &   B    & No $z$\tablenotemark{a}  \\
\tableline

DEEP total   &      813 &  619  &  52  &  142   \\
\x Galaxies  &      620 &  572  &  48  &    -   \\
\x Stars     &       51 &   47  &   4  &    -   \\
\x No redshift ID & 142 &    -  &   -  &  142   \\
\tableline
CFRS+others  &       47 &   36  &  11  &    -   \\
\x Galaxies  &       38 &   28  &  10  &    -   \\
\x Stars     &        9 &    8  &   1  &    -   \\
\tableline
DEEP+CFRS+others &      &       &      &       \\
\x Galaxies      &  658 &  600  &  58  &    -    \\
\x Stars         &   60 &   55  &   5  &    -    \\
\tableline
\end{tabular}
\tablenotetext{a}{``No $z$'' is coded as F in Table \ref{table-catalog}.}
\end{center}
\end{table}

\begin{figure}[ht]
\includegraphics[width=3.5truein]{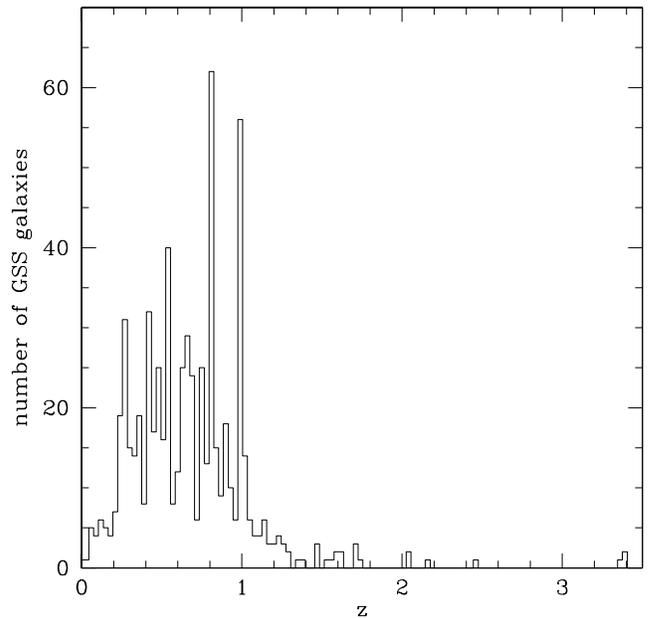}
\caption{Redshift histogram including quality A+B redshifts for the 
GSS sample, including DEEP1 and other sources.  
Sharp spikes indicate large-scale structure features
permeating the volume.  The sharp decline in the number of
galaxies near $z\sim1.05$ coincides with the 
movement of [O II] 3727 into the bright atmospheric OH band at 
7700 \AA.}
\label{fig-zhist}
\end{figure}

Figure \ref{fig-zhist} shows the final DEEP1/GSS galaxy redshift 
histogram, including 620 DEEP redshifts and 38 redshifts from other 
sources.
The median $z$ is 0.651,
and the number of galaxies drops off rapidly above $z=1.05$.  
This decline is probably due to the redshifting of
[O II] 3727 into the bright atmospheric OH band at 
7700 \AA, where it can be masked by poor sky subtraction.
The redshift completeness is discussed in the next section. 
A few objects are identified at very 
high redshift by their UV interstellar absorption lines
or broad AGN emission.  One object at $z=3.40$ is the source
in the quadruple gravitational lens galaxy 093\_2470, identified 
through Ly$\alpha$ emission.

\subsection{Completeness}
\label{sec-zfail}

The completeness of the DEEP1 spectroscopic survey is defined as the
fraction of observed targets successfully identified as galaxies or
stars with quality A or B.  Here, observed targets include all objects
for which a usable spectrum was extracted, including alignment stars
but not including those objects that fell off the slit, off the edge
of the CCD, or were otherwise corrupted.  Since there was zero
possibility to obtain redshifts for these, it is as if they had never 
been observed.  In this section, we discuss only DEEP1 targets and
omit redshifts from other surveys.

\begin{figure*}[ht]
\centerline{
\includegraphics[width=5.0truein]{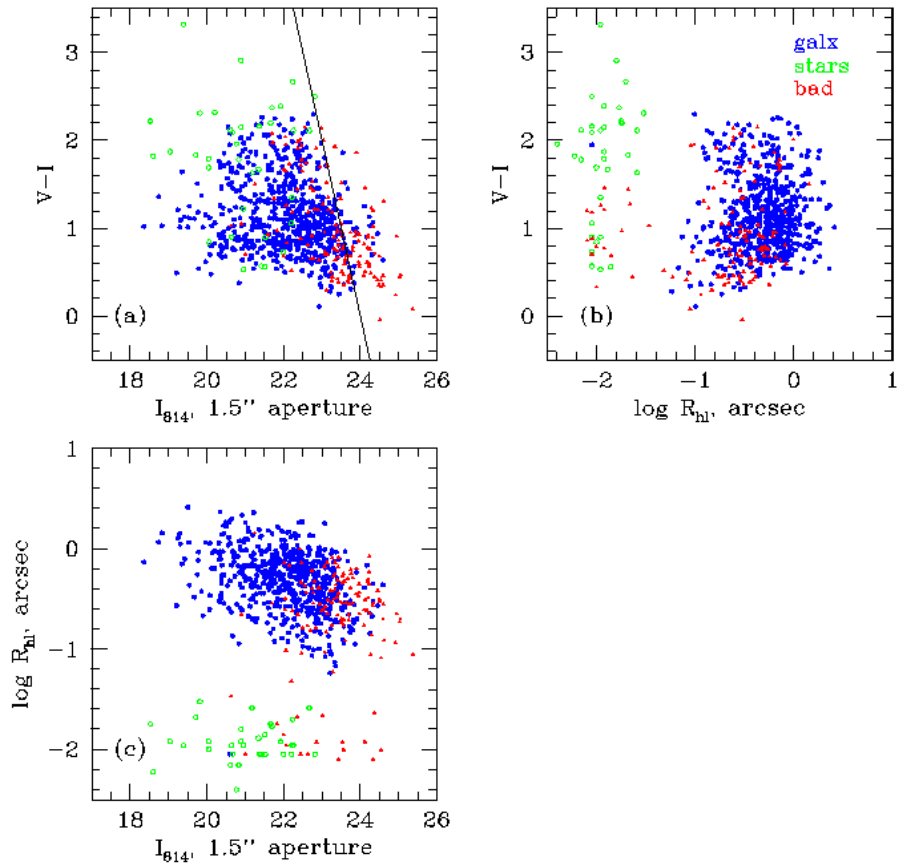}
 }
\caption{Redshift success versus apparent magnitude, color, and
half-light radius for all spectroscopic targets.  Blue filled circles
= galaxies, green open circles = stars, red triangle = no redshift.
The nominal selection limit of the survey, $(V+I)/2>24$, is 
shown by the diagonal line in panel (a).
Objects with failed $z$'s are more frequently faint,
blue, and small, but most are not stars, which are pointlike.
Followup spectroscopy by DEEP2 and other surveys shows
that most faint blue objects lie beyond the useful range
of the present survey, namely $z \sim 1.05$, where OH interferes
with [O II] 3727 (cf. Figure \ref{fig-zhist}).
}
\label{fig-colormagrad_app}
\end{figure*}

Figure \ref{fig-colormagrad_app} shows 
the apparent color-magnitude diagram for 
galaxies, stars, and targets without redshifts.  
%This shows generally
%where redshift failures lie in color-magnitude space.  
The most striking feature of this diagram is the concentration of
failures at faint magnitudes and very blue colors ($V-I<1$).
Such blue galaxies are expected to have strong emission, so
it is unlikely that they were missed on account of weak
spectral features.  We suspect that they simply lie beyond
the effective redshift range of the survey, $z \sim 1.05$, set by bright
OH at 7650 \AA.  This is confirmed by
subsequent DEEP2 observations, which are taken at higher spectral
resolution, are less contaminated by OH emission,
and reveal large numbers of galaxies beyond $z = 1.05$ out 
to the DEEP2 redshift limit $z = 1.4$; many of these occupy the
region where DEEP1 failures lie.
Further information on blue redshift failures is provided by
measurements probing the ``redshift desert'' by 
Steidel \etal\ (2004), who show that many galaxies 
successfully recovered in the range
$z = 1.5-2.5$ display colors similar to the blue failures
here.  Some of these are actual followup observations of DEEP2 failures 
by that group, showing most of them to lie beyond $z=1.4$.  In totality,
this information is powerful
evidence that the bulk of blue redshift failures in DEEP1 lie
beyond $z = 1.05$.

The situation is different for red failures with $V-I > 1.5$,
which tend to lie further above the nominal survey magnitude limit,
as seen in Figure \ref{fig-colormagrad_app}.
At $22.5<(V+I)/2<23.5$, the failure
rate is higher for red objects than for
blue objects.  This is plausible because 
it is harder to identify a redshift for faint red objects with 
weak or no emission lines.

\begin{figure}[ht]
\includegraphics[width=3.5truein]{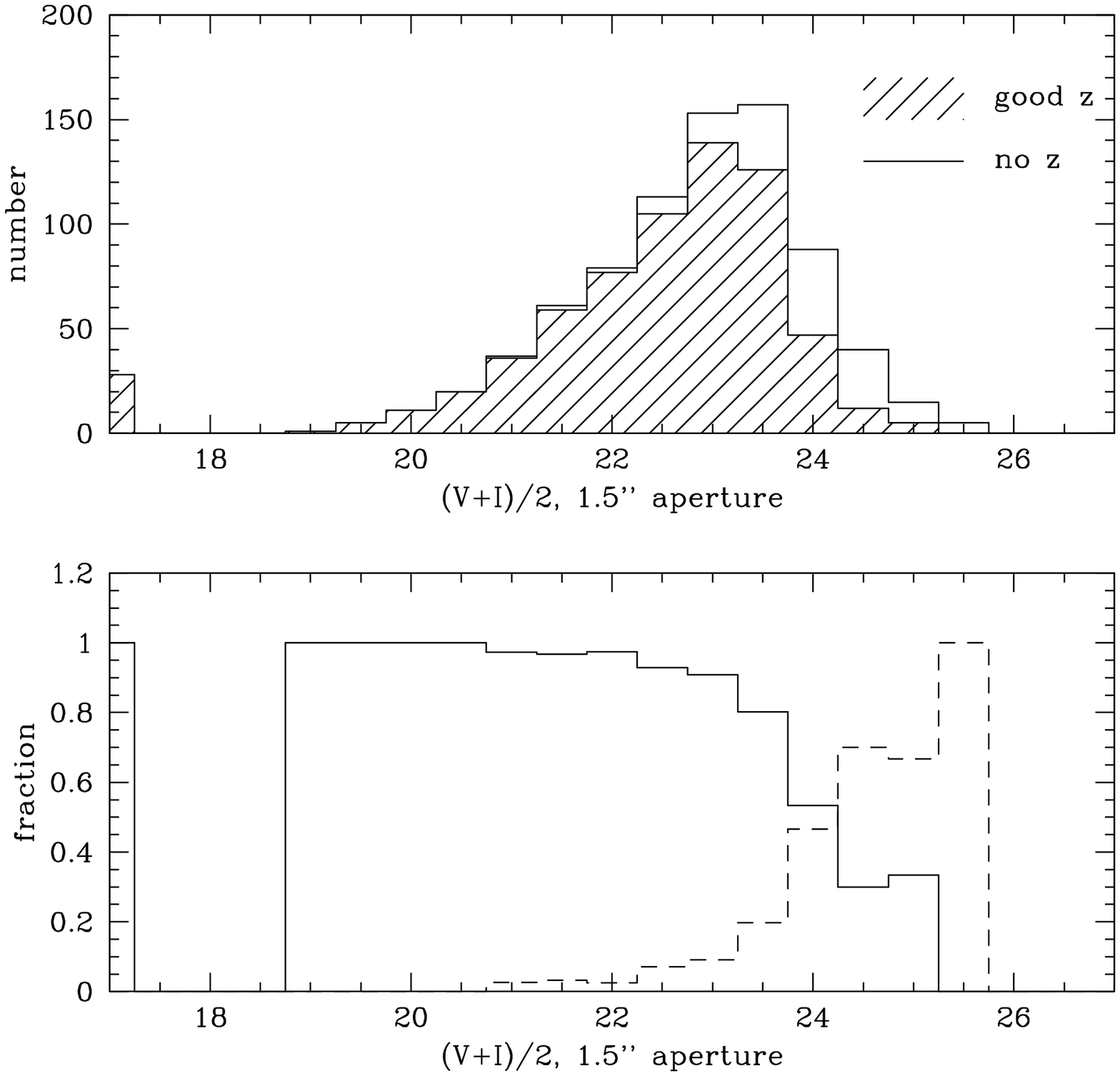}
\caption{The dependence of redshift success
on $(V+I)/2$, by number (top) and by fraction 
(bottom, success: solid, no redshift: dashed). 
The half-power success point is $(V+I)/2 = 24$,
which may be taken as the effective magnitude limit of
the survey.}
\label{fig-completei}
\end{figure}

\begin{figure}[ht]
\includegraphics[width=3.5truein]{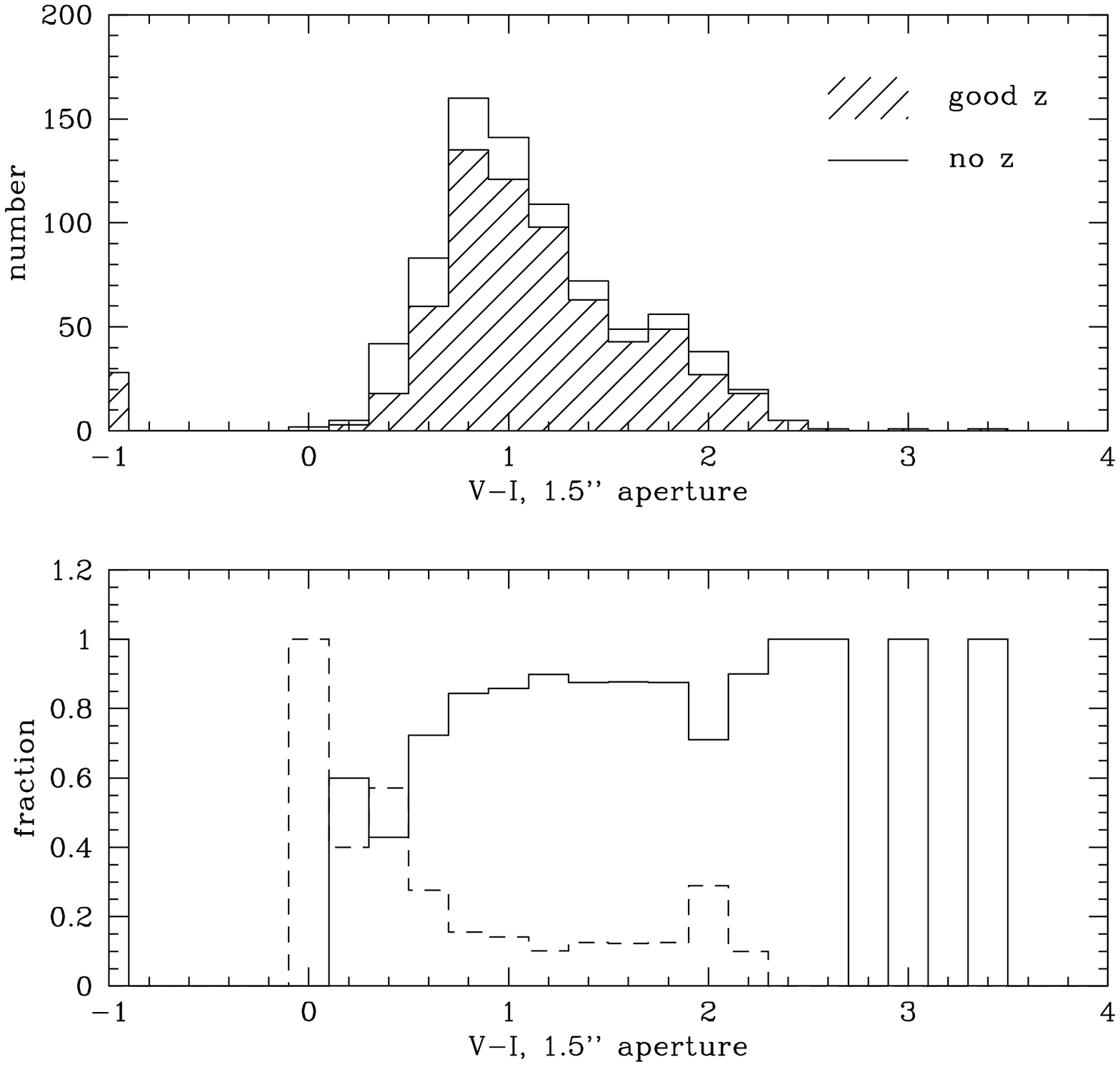}
\caption{The dependence of redshift success
on $V-I$ color, by number (top) and by fraction
(bottom, success: solid, no redshift: dashed).  
Figures \ref{fig-colormagrad_app} and \ref{fig-completei}
combined with this one show that redshift failures are primarily blue 
and faint.  Additional spectroscopy indicates that most of these are 
objects beyond our effective redshift limit $z\sim1.05$.}
\label{fig-completevi}
\end{figure}

Figures \ref{fig-completei} and \ref{fig-completevi}
quantify the information in Figure \ref{fig-colormagrad_app}.  
Figure \ref{fig-completei} plots
the number and fraction of targets and
quality A+B identified objects as a function of $(V+I)/2$ magnitude.
The completeness fraction declines only slowly with magnitude,
remaining at 80\% at $(V+I)/2=23$, but drops to 50\% in the $(V+I)/2=24$
bin.  Only a few objects are identified at $(V+I)/2>24$.  Thus the limiting
magnitude of the survey is effectively $(V+I)/2 = 24$.

Figure \ref{fig-completevi} plots the completeness fraction against
color.  Incompleteness increases strongly for galaxies bluer than
$V-I=1.0$, and for the bluest objects, it is very high.  As noted, we
believe these objects to be at $z \gtrsim 1.05$.

\begin{figure}[ht]
\includegraphics[width=3.5truein]{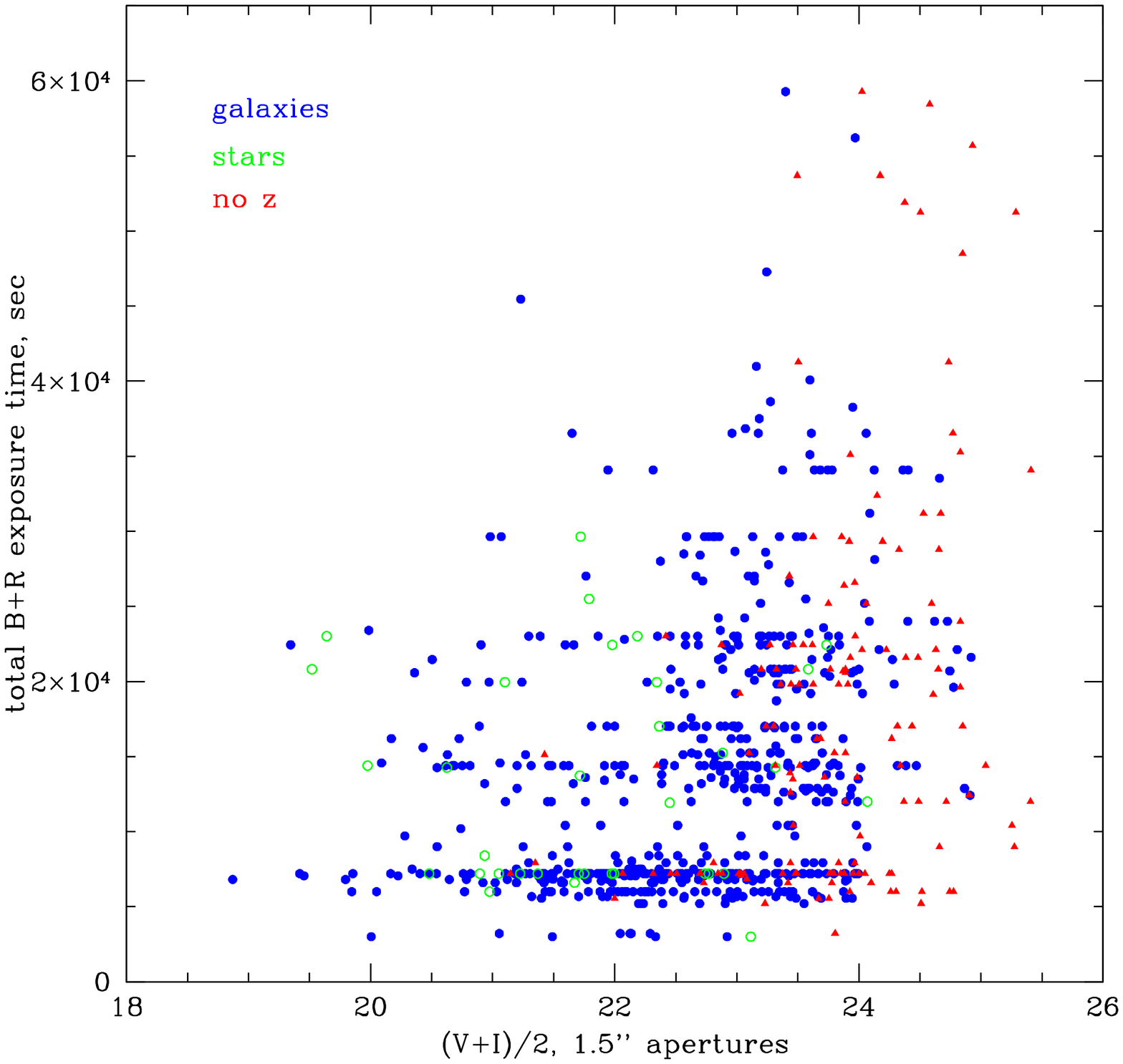}
\caption{Redshift success in $(V+I)/2$ magnitude and exposure time space.
The total exposure time (blue+red wavelength settings) versus $I$ magnitude,
with points coded for galaxies = blue circles, stars = green open circles, 
and no redshift identified = red triangles.}
\label{fig-exptime}
\end{figure}

Figure \ref{fig-exptime} shows the distribution of targets in
magnitude and exposure time with points coded as confirmed galaxies,
confirmed stars, or no redshift.  
This figure suggests
that increased exposure time on the faintest objects produces only a
marginal improvement in the success rate.  This is again understandable if
most of the failed objects are 
are outside the effective wavelength range.
This argument applies to the blue galaxies, but one would 
expect that red failures might yield more frequently to longer
exposure times.  Unfortunately, the distribution of exposure
times for red galaxies does not permit a test of this.

Figure \ref{fig-colormagrad_app} provides additional information
on the apparent sizes of objects.  In other redshift surveys,
with objects selected from ground-based images, the
separation between stars and galaxies is sometimes blurry.
The excellent resolution of HST for all objects
here removes that problem, and
Figure \ref{fig-colormagrad_app} shows a clean separation
between pointlike objects and resolved galaxies.  The separation
is especially clean because the HST PSF is accounted for.

Of pointlike objects with redshifts, only two
prove to be non-stellar: a QSO and a subcomponent of
an interacting galaxy.  Thus, most of the
pointlike objects without successful redshifts
are presumably stars.  Most
stars at these faint magnitudes are cool dwarfs with broad
absorption bands, or hotter stars with nearly featureless spectra,
occasionally showing H$\alpha$ or Ca II absorption.  We are
generally able to positively identify red, cool stars with 
$I\lesssim22$ from their absorption bands.  Fainter than $I=22$, the
continuum is too weak to make positive identifications.
The hotter stars are more
difficult since their spectra have few or no features. 
Figure \ref{fig-colormagrad_app} shows a number of unidentified
pointlike objects with $0.5<V-I<1.5$ that are presumably
hot stars.  There appears to be no substantial population
of intrinsically small, pointlike field galaxies.\footnote{
The ultra-compact dwarfs found in the Fornax cluster 
(Philipps \etal\ 2001) are much fainter than galaxies in our sample,
and probably a cluster population.}  Surveys which 
weed out pointlike objects will miss QSOs, but will not miss a 
significant number of galaxies if their star/galaxy separation 
is good.  

\section{Galaxy colors}
\label{sec-colors}

Restframe colors and magnitudes are essential to study the
evolution of galaxy properties.  For galaxies
with redshifts, observed
HST $V_{606} - I_{814}$ color from Table \ref{table-catalog}
is used to infer both restframe
$U-B$ color and K-correction ($K_{IB}$) from
observed $I$ to rest $M_B$.  
To obtain K-corrections and restframe
colors, we use a set of 34 galaxy UV-optical spectral energy
distributions selected 
from the atlases of Calzetti \etal\ (1994) and Kinney \etal\ (1996).
The SEDs are those of ellipticals, spirals, and starbusts
with continuous wavelength coverage; a few abnormal spectra were
excluded.

The first step in the K-correction is to synthesize a restframe $U-B$ color 
for each of the atlas SEDs using the system response curves for each
filter and the zeropoints of the UB Vega
system (Fukugita \etal\ 1995).  
Next, for each DEEP
galaxy we redshift each atlas SED to the redshift of the galaxy and 
convolve the $f_\lambda$ SEDs with the response curves to
synthesize an observed HST $V-I$ color and K-correction $K_{IB}$
from observed $I$ to restframe $M_B$.  We then fit low-order polynomials
to the resultant atlas $U-B$ and $K_{IB}$ values as a function of 
synthesized $V-I$.  Entering these fits with
the observed $V-I$ color of the DEEP1 galaxy yields $K_{IB}$ and  
$U-B$.  A few very blue galaxies are bluer than any of the SEDs and
are truncated to $U-B=-0.6$ to avoid extrapolation.
The procedure is described further in Willmer \etal\ (2004, in preparation) 
and is similar to that used in Gebhardt \etal\ (2003) except that here we
fit a polynomial over $V-I$ at each galaxy redshift rather than over 
$V-I$ and $z$ simultaneously.
The resultant values of $U-B$ and $M_B$ are given in Table 
\ref{table-catalog}.\footnote{
The atlas SEDs are corrected for Galactic extinction but not
internal extinction.
% so the procedure yields atlas $V-I$ and restframe $U-B$ values
% that are corrected for Galactic but not internal extinction.
The observed $V$ and $I$ magnitudes of GSS galaxies from Simard
\etal\ (2002) have not been corrected for either source of 
extinction.  Galactic extinction in the GSS is 0.025 mag in $V$ and
0.014 in $I$ (Schlegel, Finkbeiner \& Davis 1998).  We have ignored
this 0.01 mag mismatch in $V-I$ in what follows.}

Error estimates in restframe $U-B$ prove to be important,
especially for red galaxies.
The error from photon statistics in raw $V-I$ is relatively small.  
The Monte Carlo error
estimation of Simard \etal\ (2002) models photon
statistics and errors from the photometric model fits, yielding median
errors in $V-I$ of 0.055 mag for this sample.
We have also compared the model colors
to aperture colors measured through a 1.5\arcsec\
aperture, and find that the rms scatter between the two colors
is only 0.085 mag, which is consistent.
The transformation from $V-I$ to $U-B$ mutiplies
the apparent color error by a factor dependent on redshift, 
from $\sim 1.0$ at $z\sim0.3$
to 0.6 at $z\sim0.9$, yielding net random errors in $U-B$ of 
0.03-0.04 mag from $z=0.4$ to 1.0.

The remaining sources of error are systematic errors in the transformation
from $V-I$ to $U-B$.  At $z=0.8$, the $V$ and $I$ bandpasses
correspond very closely to restframe $U$ and $B$, and the color
K-correction is small.  At redshifts other than $z=0.8$, the color
does not translate so closely, but our K-correction procedure
yields accurate restframe colors for $0.4<z<1.2$; nearer
than that, the correction is less reliable.  
We have checked that the procedure introduces minimal scatter
using methods to be explained in Section \ref{sec-scatter},
but it is possible that there are still small zeropoint
shifts as a function of redshift.  We also
have completely independent ground-based photometry for all galaxies
from the Canada-France-Hawaii Telescope in $B$, $R$, and $I$ filters
(Kaiser \etal\ in preparation), and have used these to
compute an independent restframe $U-B$ color.  For galaxies at 
$z>0.4$, the CFHT-derived $U-B$ values agree well with the present ones
with an rms scatter of 0.12 mag and a zeropoint offset of 0.02 mag, the
HST colors being redder.  

A final source of systematic error is the difference between colors 
from an energy-weighted convolution of the $f_\lambda$ SED with 
the response curve, as described above, and from a photon-weighted
convolution.  In principle, since no color term was applied to
the HST magnitude solution (Simard \etal\ 2002), the photon-weighted
convolution is correct when the HST zeropoints are determined for
an object the color of Vega.  In practice, photon-weighting makes the
inferred $U-B$ colors redder, but the difference is only a mean
of 0.02 mag and RMS of 0.014 mag, and we have not applied this offset
to the tabulated values.
To sum up, we believe that the typical
random error of $U-B$ is 0.03-0.04 mag in the critical redshift range
$0.4<z<1.0$, with systematic errors in zeropoint at $z=0.8$
of $\sim 0.02-0.03$ mag.

\section{Results}

\subsection{Redshift Distribution}

\begin{figure}[ht]
\includegraphics[width=3.5truein]{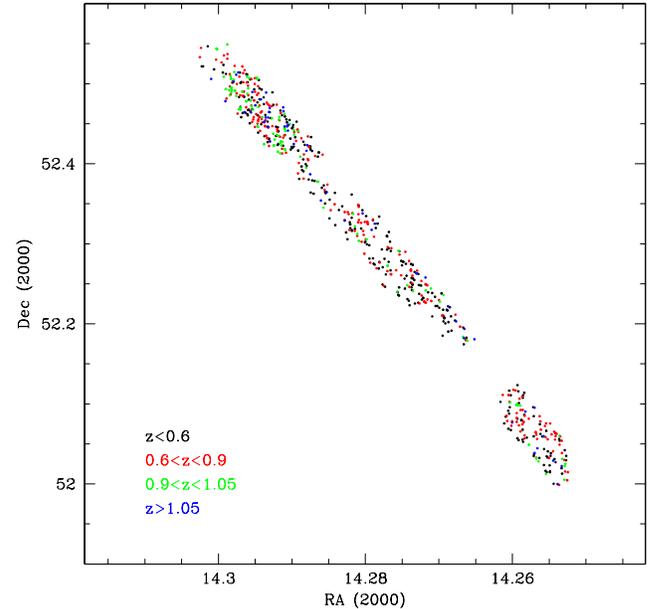}
\caption{Distribution of DEEP 1 GSS galaxies on the sky.
Galaxies are color coded by redshift range: black: $z<0.6$;
red:$0.6<z<0.9$; green:$0.9<z<1.05$; blue:$z>1.05$.}
\label{fig-skydist}
\end{figure}

The distribution on the sky of DEEP1/GSS galaxies with redshifts is
shown in Figure \ref{fig-skydist}.  The spectroscopy samples an area 
approximately 3\arcmin\ $\times$ 40\arcmin.  Here and subsequently,
we include the 38 galaxy redshifts from sources outside DEEP1,
located primarily at the northeast end of the Groth Strip.
Some areas of the Groth Strip are more
densely covered than others, and there are patches without 
any coverage at all.  The limited area and non-uniform
sampling preclude quantitative 
statements about the 3-D galaxy distribution
or local galaxy overdensities.
These statistics are much better probed in the ongoing
DEEP 2 survey, which has larger area and
statistically uniform sampling (e.g., Coil \etal\ 2003).

\begin{figure*}[ht]
\centerline{
\includegraphics[angle=-90,width=6.0truein]{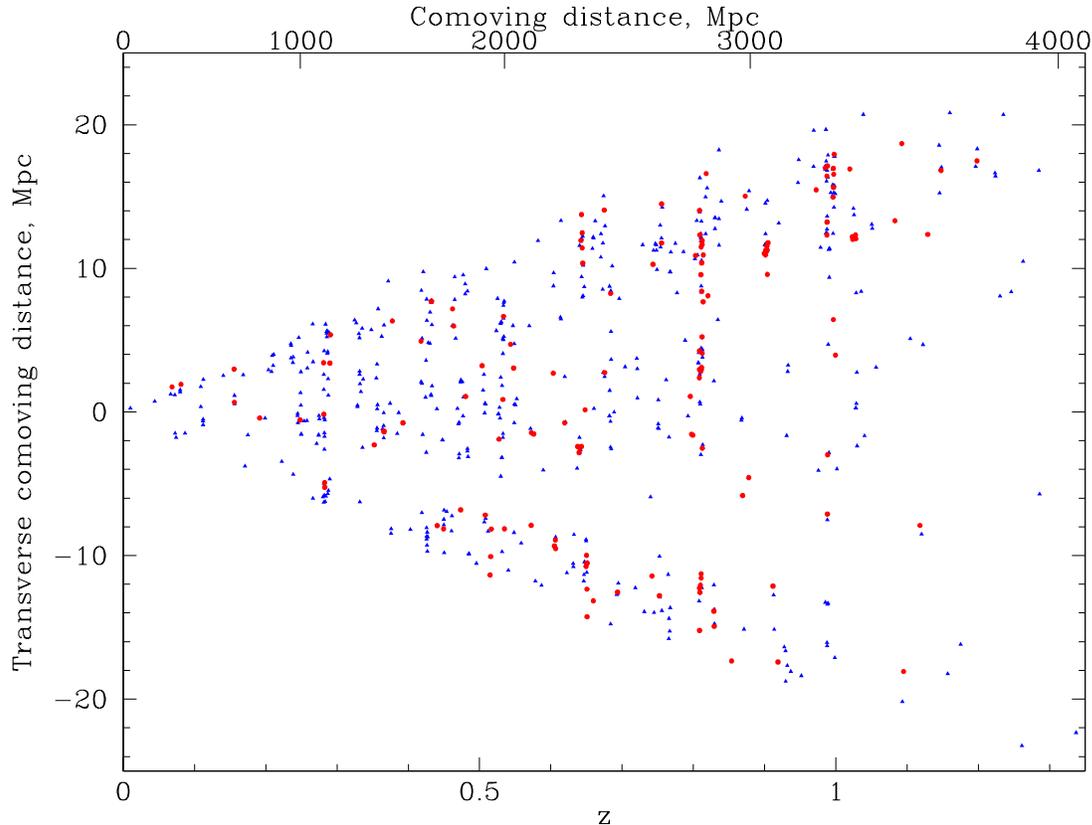}
}
\caption{``Pie'' diagram of galaxies in the Groth Strip.  
The transverse spatial dimension is stretched to show
points more clearly.  
Large-scale structures of voids and walls persist to $z\sim1$.
Blue galaxies are plotted as blue triangles,
and red galaxies as red filled circles; red galaxies appear to populate
denser regions, as demonstrated quantitatively in the clustering analysis
of DEEP2 data by Coil \etal\ (2003).}
\label{fig-piediag}
\end{figure*}

The ``pie diagram'' in Figure \ref{fig-piediag} plots
transverse distance of each galaxy in comoving
Mpc along the slice versus redshift. 
``Walls'' and filaments due to large-scale 
structure cross the volume, with the most prominent
ones at $z=0.28, 0.81, 0.99$.  The transverse distance is 
stretched to enhance the visibility of points in dense regions.

Galaxies in
the $z=0.99$ wall stretch over 37 comoving Mpc,
demonstrating that large-scale structure is already 
prominent at $z\sim 1$.  The $z=0.99$ feature was already 
noted as a redshift peak in the CFRS survey
by Le F\`evre \etal\ (1994, 1996)
and in early DEEP1 data by Koo \etal\ (1996).  
Filamentary/wall structures on this and even larger scales
are prominent in local surveys, as shown in 
the Las Campanas Redshift Survey (Doroshkevich 
\etal\ 1996), the 2dF Survey (Colless \etal\ 2001), and 
the SDSS Survey (Zehavi \etal\ 2002), and strong redshift
peaks were noted in early high-$z$ field galaxy
redshift surveys and ascribed to large-scale
structure (e.g., Broadhurst \etal\ 1990; Willmer \etal\ 1994;
Le F\'evre \etal\ 1994; Cohen \etal\ 1996; Connolly \etal\ 1996).

It is therefore no surprise, given our knowledge of how structure
forms and the nature of galaxy biasing,
to find large, well developed structures already outlined by galaxies
at $z \sim 1$.  Figure \ref{fig-piediag} is one of the first 
cuts through the Universe to show a full 2-D view 
of this structure to high redshift, other ones being 
early DEEP2 data by Coil \etal\ (2003) and early VIMOS VLT Deep Survey
(VVDS) data by Le F\`evre \etal\ (2004).
The three pictures are qualitatively very similar.

\subsection{Color bimodality}

\begin{figure*}[ht]
\centerline{
\includegraphics[width=5.5truein]{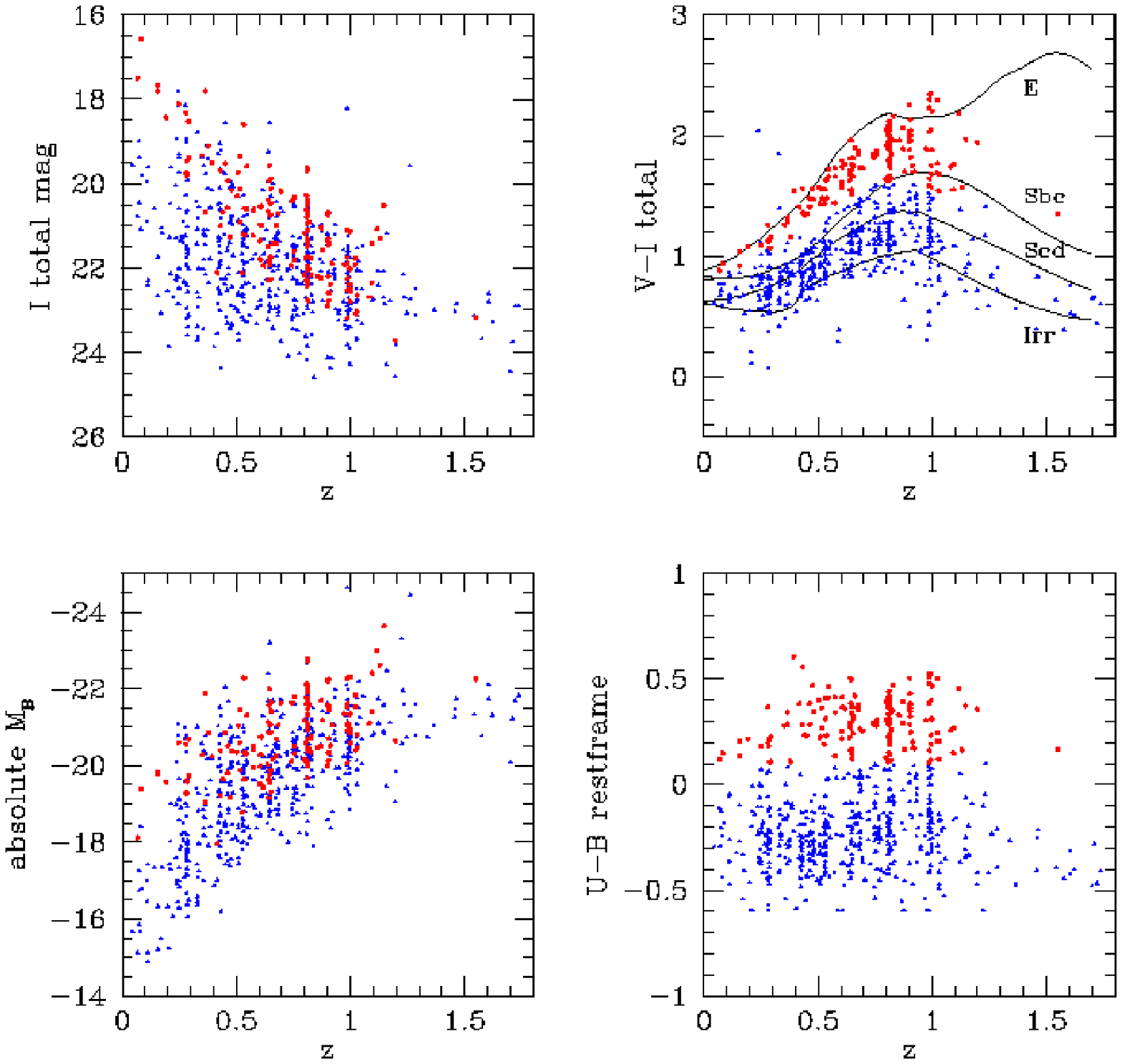}
 }
\caption{Galaxy magnitudes and colors versus
redshift.  Color bimodality is clearly seen in both apparent $V-I$
and restframe $U-B$.  Galaxies are plotted as red circles or blue triangles
according to whether $U-B$ is greater or less than 0.1.  Color tracks
of non-evolving local-galaxy SEDs from Coleman, Wu, and
Weedman (1980) are shown in the upper right panel.
No single CWW track exactly follows
the valley between red and blue galaxies.}
\label{fig-zmagcolor}
\end{figure*}

\begin{figure}[ht]
\includegraphics[width=3.5truein]{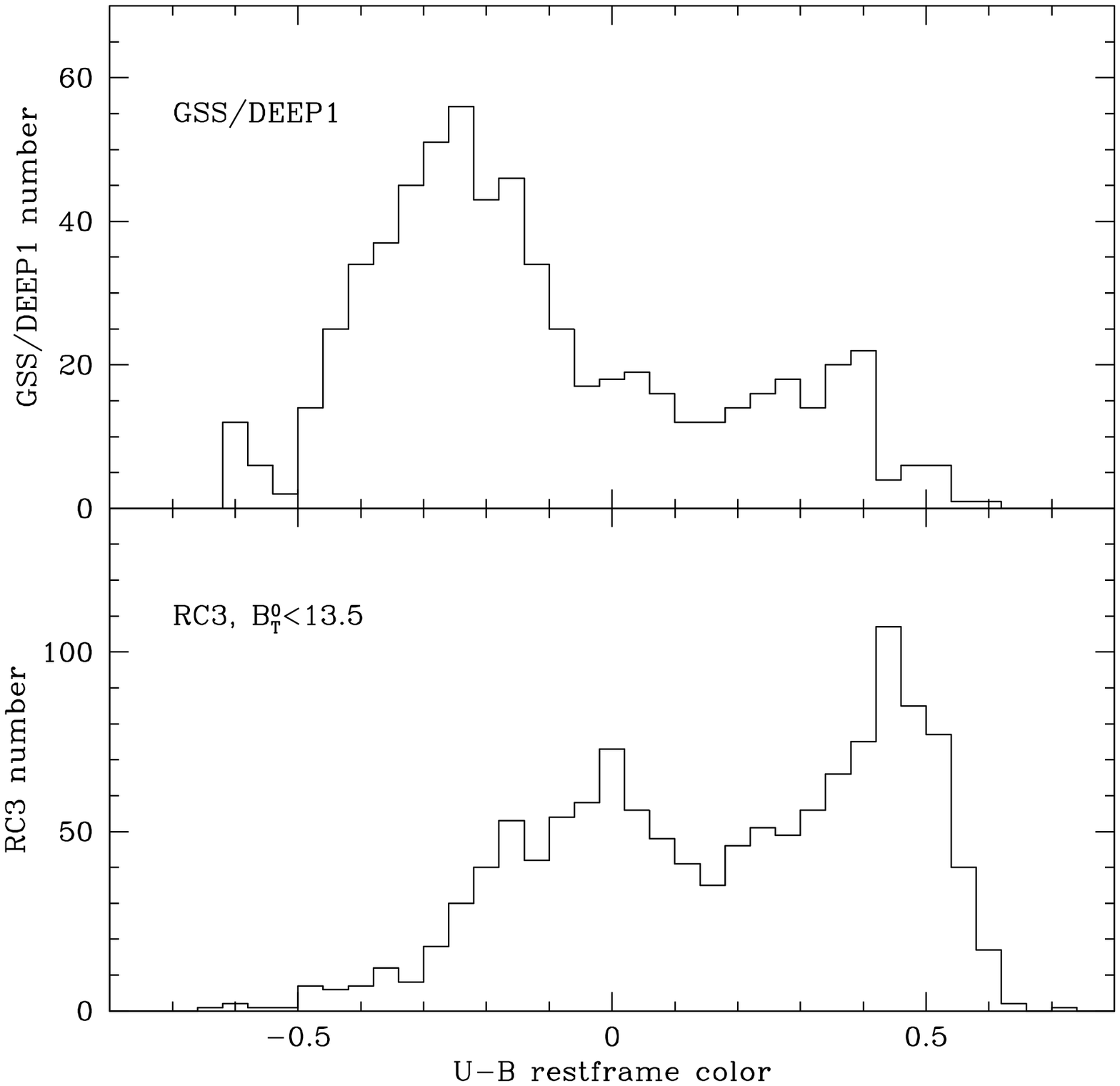}
\caption{
Upper panel: Total restframe $U-B$ color distribution for DEEP 1 galaxies.
Lower panel: Total $U-B_T$ color distribution for galaxies from the RC3 catalog
with $B_T^0<13.5$.}
\label{fig-ubhist}
\end{figure}

Figure \ref{fig-zmagcolor} gives an overview
of magnitudes and colors (both observed and restframe)
versus redshift.  
The most striking feature
is the bimodality in restframe $U-B$ (panel d).
Galaxies are divided into red and blue 
populations, and this division persists to $z=1$.
The division is at approximately $U-B = 0.10$, 
which has been used to color-code the points in 
Figures \ref{fig-zmagcolor} and \ref{fig-piediag}.  
Figure \ref{fig-ubhist} shows
the bimodal distribution of restframe $U-B$ color in DEEP1
and, for comparison, the colors of a local sample from the
RC3 (de Vaucouleurs \etal\ 1991).  We expand on this comparison
in Section \ref{sec-localcomp}.

The existence of restframe color bimodality in GSS data
confirms a similar hint noted by Im \etal\ (2002)
for DEEP1 data based on apparent color,
and is consistent with the definitive
discovery of color bimodality for distant galaxies
by COMBO-17 (B03) based on a much larger sample.  The same 
effect is seen in DEEP2 data using principal component 
analysis of galaxy spectra (Madgwick \etal\ 2004).
Bimodality has been much discussed for local galaxies 
by the SDSS collaboration (e.g., Strateva \etal\ 2001; Blanton
\etal\ 2003; Hogg \etal\ 2003), and for 2dF data (Madgwick \etal\ 2002)
although it turns out to have been
clearly present in RC3 data (de Vaucouleurs \etal\ 1991) all along
(see below and Figure 8 of Takamiya \etal\ 1995).  
Roughly speaking, red galaxies are consistent
with passive populations and little star-formation, or with 
dust reddening (see Section \ref{sec-emission}), while blue
galaxies are star-forming.  A galaxy which stops
forming stars will fade and redden across the blue/red divide
in a few Gyr.  As emphasized by Kauffmann \etal\ (2003),
bimodality for local galaxies is not restricted to color alone
but extends to many other galaxy properties 
such as luminosity, mass, concentration index, 
star-formation history, and environment, which are all correlated
with color.  

The existence of color bimodality is thus a fundamental
property of the galaxy population, and
its presence out to $z\sim1$ means that
it was established fairly early in the evolution of galaxies.  
An important task
is to ascertain whether the other properties 
that correlate with bimodality locally (such as luminosity,
mass, etc.) also do so for distant galaxies.  We 
explore such correlations in the remaining sections of this paper.

In addition to being of theoretical interest, 
bimodality is of practical significance because
it means that distant galaxies can be accurately sorted into
groups using an easily measured parameter, color.
This is needed because it is obvious that
different classes of galaxies have had different evolutionary
histories, and it is even possible that some
galaxies have evolved from one type into another.  Sorting,
counting, and tracking by type is thus essential.  In previous
surveys, this has often been done by dividing galaxies
according to the non-evolving SEDs of Coleman, Wu, \&
Weedman (1980), as illustrated in panel (b) of
Figure \ref{fig-zmagcolor}.  The Sbc SED, second from the top,
has been used to divide
red and blue galaxies (as in the CFRS, Lilly \etal\ 1995b).
This SED does not quite follow the
valley between red and blue galaxies.  Dividing
galaxies at this SED can therefore mix the two types; sorting by any
non-evolving SED or color can introduce 
redshift-dependent errors in the counts if the location of the 
valley evolves.  It is safer to have an index of red/blue that
evolves along with the galaxy population itself, which
color provides, if the sample is large enough.  In this 
paper we will generally divide red and blue at restframe
$U-B=0.1$, which is applicable to $0.4<z<1.0$ where the majority
of our sample lies, but not in the local universe.  Samples 
with very large numbers such as COMBO-17 (B03) or DEEP2 allow a
finer determination of the evolving division.

Interestingly, a division by color for high-redshift galaxies is not seen
in CFRS data (Crampton \etal\ 1995).  Possibly the
bimodality was washed out by the relatively larger color errors 
in the CFRS 3\arcsec\ ground-based aperture ($\sim 0.15$ mag, 
Le F\`evre \etal\ 1994).
This highlights the desirability of obtaining precise colors 
in studies of distant galaxy evolution.
The CFGRS approached classification for high-redshift galaxies
somewhat differently, using spectral typing into 3 classes rather 
than restframe color; there is a hint of a bimodal distribution in 
spectral index in Figure 8(a) of Cohen (2001).
The CNOC2 survey divided galaxies into 3 classes based on 
SED types derived from CWW SEDs and fit to the apparent colors, a scheme
that is closely related to restframe color 
(Lin \etal\ 1999).  Their sample from $0.12<z<0.55$
shows a distinct bimodality in SED type, but the 3 galaxy classes
were not chosen to follow the bimodality.

\subsection{Color-magnitude diagram}

\begin{figure*}[ht]
\centerline{
\includegraphics[angle=-90,width=7.5truein]{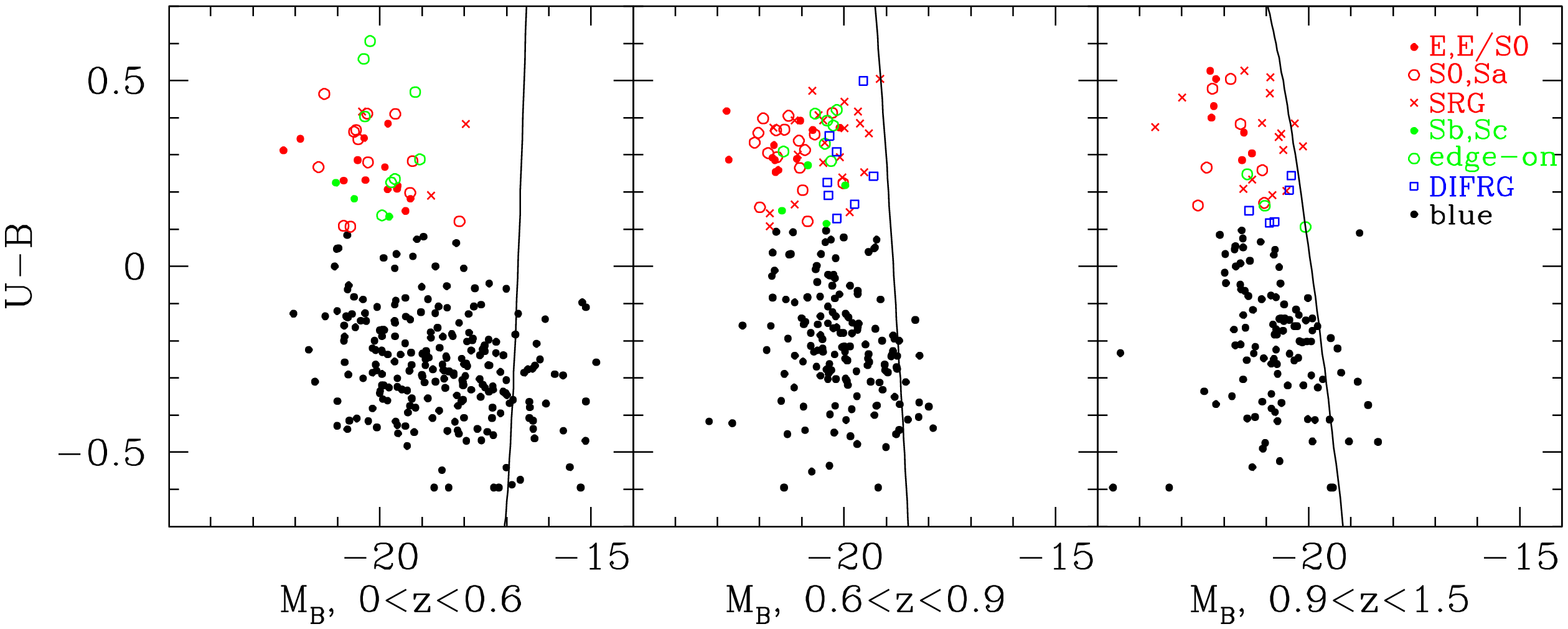}
}
\caption{Restframe color-magnitude diagram for Groth
Strip galaxies in three redshift ranges.  The dividing line for 
red galaxies used here is $U-B$ = 0.1.
Points for red galaxies are color-coded
by morphological type: red filled circle = E; red open circle = E/S0-Sa;
red x = small red spheroidal galaxy;
green filled circle = Sb-Sc; green open circle = edge-on disk; 
blue open square = diffuse red galaxy.
Galaxies in the blue population (below $U-B = 0.1$)
are not classified.  The magnitude limit of $(V+I)/2=24$
for the median redshift in each bin, $z=0.4,0.75,1.0$, 
is shown as the near-vertical line in each panel.
}
\label{fig-colormag}
\end{figure*}

Figure \ref{fig-colormag} shows the restframe color-luminosity
diagram of the DEEP1 galaxies.  
The bimodal division in color also 
corresponds to a difference in the distribution of luminosities:
red galaxies are slightly more luminous in restframe $B$ than the 
brightest blue galaxies.
The distribution in color and magnitude is
qualitatively consistent with the distributions
seen in the SDSS sample locally (Blanton \etal\ 2003;
Hogg \etal\ 2003) and with the CM diagrams
of B03.  Luminosity functions computed by color
show a systematic dependence of $L^*$ on
color for local galaxies, with $L_B^*$ being more than 1 magnitude
brighter for the reddest local galaxies compared
to the bluest in SDSS (Blanton \etal\ 2001).
Similar behavior is seen at high redshift in the COMBO-17
sample (Wolf \etal\ 2003),
and in both DEEP1/GSS and early
DEEP2 data by Willmer \etal\ (2004, in preparation).
Figure \ref{fig-colormag} also codes red galaxies by morphological
type, discussed in Section \ref{sec-morphology}.

Figure \ref{fig-colormag} shows the changing effect of
the magnitude selection limit $(V+I)/2=24$ as a function of redshift.
The near-vertical lines in each panel indicate the selection limit 
at the median redshift for that bin, $z=0.4$, 0.75, 1.0 respectively.
At low redshifts, the selection, which is effectively $R$ band,
favors intrinsically red galaxies slightly.  At high redshifts, the 
selection moves to restframe blue or near-UV wavelengths, and 
intrinsically red galaxies are disfavored.  Beyond $z=1.1$, there are 
few red galaxies in the sample, simply because the steep drop in their
luminosity function at $L^*$ means few exist which are bright enough
to be selected.  This effect is also seen in the $M_B(z)$ panel 
of Figure \ref{fig-zmagcolor}.  

However, at low redshift, we find few {\it faint} red galaxies, 
even though they are not selected against; measurements 
of the luminosity function by color are discussed further
in Willmer \etal\ (2004, in preparation).  
Some low-redshift field-galaxy surveys 
find that the luminosity function of red galaxies falls
at faint magnitudes (e.g., SDSS, Blanton \etal\ 2001).  Our data
appear to agree, though
care must be taken in interpreting the LF falloff, since the 
color-magnitude relation slope for red galaxies means that 
a division at fixed restframe color can produce a red LF
which falls at the faint end.

\begin{figure}[hbt]
\includegraphics[width=3.5truein]{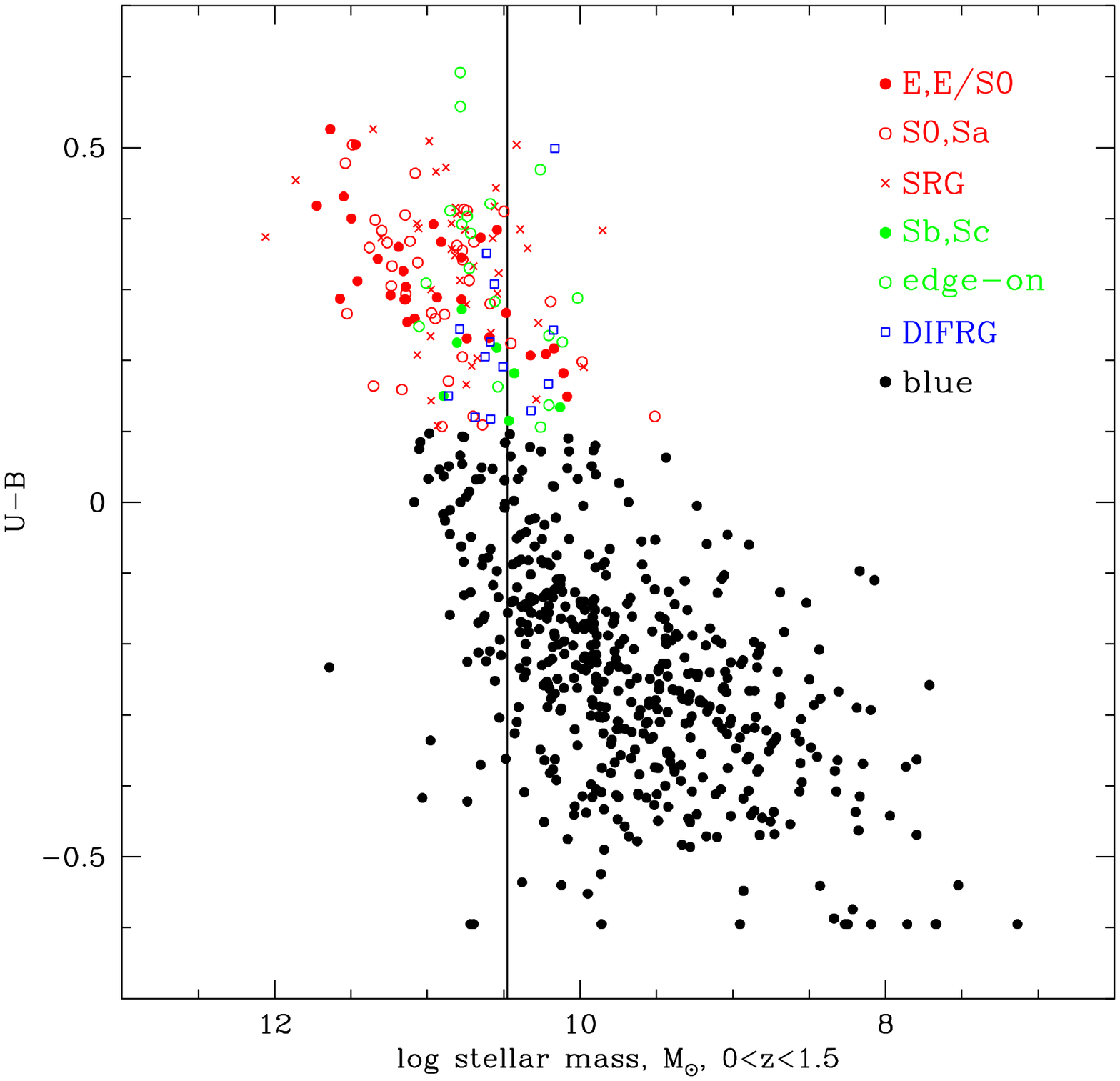}
\caption{Restframe color versus approximate stellar mass.  Mass is 
derived from magnitude and color following the prescription
of Bell \etal\ (2003c).  
Red galaxies are coded for morphological type as
in Figure \ref{fig-colormag}.  The vertical line is the break 
at $3 \times 10^{10} M_\odot$ in properties of local galaxies
found by Kauffmann \etal\ (2003).
In our sample, red galaxies to the right of this line are potentially
lost at high $z$ due to magnitude selection effects. 
}
\label{fig-colormass}
\end{figure}

The mass difference between red and blue galaxies is minimized in
Figure \ref{fig-colormag} by use of blue magnitude $M_B$, 
which penalizes red galaxies.  Figure
\ref{fig-colormass} replots the CM diagram as a
function of approximate stellar mass using 
$M_B/L$ values as a function of $B-V$ (transformed
to $U-B$) from Bell \etal\ (2003c).  A downward
correction of 0.15 dex has been applied to convert
to a Kroupa IMF (Kroupa, Tout, \& Gilmore 1993),
as advised by Bell \etal\ (2003c).  
The true mass variation with color is now more apparent.  
By eye, the division between red and blue galaxies
occurs at $3 \times 10^{10}$ M$_{\solar},$ the same value
found for SDSS galaxies by Kauffmann \etal\ (2003).
%\comment{Kauffmann et al found a bimodality in $D_n(4000)$,
%not in color. - BJW}
However, the selection against red galaxies at $z \sim 1$
penalizes lower-mass red galaxies, so the lack of red galaxies
below $3 \times 10^{10}$ M$_{\solar}$ is partly a selection 
effect.

Though the dividing line between red and blue galaxies is far from
sharp in mass, it is a useful generalization that massive galaxies made
the bulk of their stars early and have now
largely shut down (Im \etal\ 2002).  The existence of color bimodality
at $z\sim1$ indicates that this process had occurred 
in many massive galaxies by that time.  The mechanism that quenched 
star-formation in the most massive galaxies is not known and is 
receiving much attention lately (e.g., Benson \etal\ 2003,
Binney 2004, Granato \etal\ 2004).  COMBO-17 (B03) estimate that the 
total stellar mass in galaxies in the red population has
{\it doubled} since $z = 1$, implying
that the quenching of a significant fraction
of the red stellar mass has occurred at surprisingly late epochs.

The brightest red galaxies are somewhat
more luminous than the brightest blue galaxies, which raises the
question of how bright red galaxies are assembled, as observed
by Bell \etal\ (B03).  Since the blue
galaxies will fade as they redden, the observed blue galaxies
cannot be the progenitors of the red galaxies, unless blue galaxies
merge at $z<1$ to build up the most massive red galaxies.
The dilemma becomes even more
apparent when color is plotted against the stellar masses of galaxies,
as in Figure \ref{fig-colormass}.  Possible alternatives are that
luminous red galaxies are assembled largely before $z=1$, that they
are built up in sites of obscured star formation, or that they are
assembled by mergers of red and/or blue galaxies.
Mergers of red galaxies are favored by B03
on the basis of their measurements of evolution in red galaxy
color and luminosity density.  Color evolution and the sloping
color-magnitude relation for red galaxies will provide
a strong constraint on models for red galaxy assembly.
The topics of quenching and mass assembly  on the red
sequence will receive intense scrutiny as surveys
like DEEP2 and VVDS gather more data.

\subsection{Morphologies of red galaxies}
\label{sec-morphology}

The WFPC2 imaging of the Groth Strip provides enough
spatial resolution for a visual, morphological classification
of most galaxies in our redshift sample.  Although morphology 
can be subjective, it is valuable for understanding 
the nature of the red/blue galaxy division and its relation
to galaxy structure.  The
red population at $z=0$ is known to consist largely of early types (E-Sa), 
while the blue population contains later-type disk galaxies
and irregulars (Sb-Irr).  This is shown below using RC3 data (Figure
\ref{fig-colormagrc3}).
The question then arises, what are the morphology and
structure of red and blue galaxies at higher redshift?
The following discussion focuses
on red galaxies; blue galaxies are deferred to a future paper.

\begin{figure*}[ht]
\centerline{
\includegraphics[angle=-90,width=6.5truein]{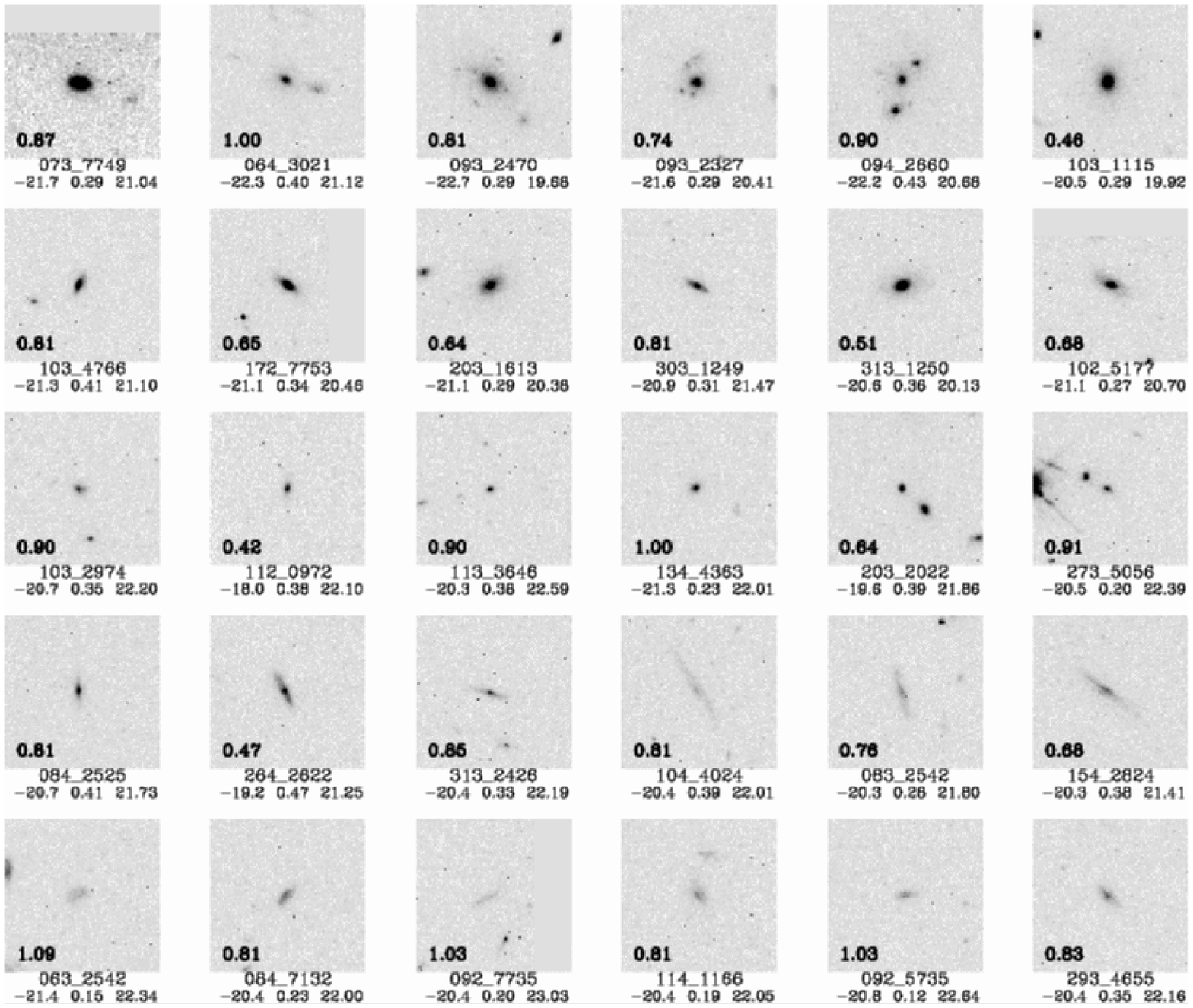}
}
\caption{
HST WFPC2 $I_{814}$ images of distant red galaxies illustrating the
three major morphological groups.  Images are 10 arcsec on a side.
The first three rows contain
spheroid-dominated red galaxies, or SRGs: the first
row shows E types, the second E/S0 through Sa, and the third row
shows small spheroidals, which are too small to determine a
detailed type.  The fourth row shows the second group, edge-on
galaxies; detailed Hubble types cannot be assigned to these
due to orientation, though some clearly have spheroids.  
The bottom row shows the third major group,
diffuse red galaxies, or DIFRGs.
The panels are labeled with galaxy redshift (in panel), GSS ID,
$M_B$, $U-B$, and $I$ magnitude.
}
\label{fig-redimages}
\end{figure*}

\begin{table}[t]
\begin{center}
\caption{Morphological types of red galaxies}
\label{table-redmorph}
\begin{tabular}{lrrrr}
\tableline\tableline
        &              \\
Type    &   Code       & $z<0.7$ & $z>0.7$ & Total   \\
\tableline

E            & -5       & 15 &  13 &  28    \\
E/S0, S0, Sa & -3,-2,0,1 & 24 &  16 &  40    \\
Small SRG    & 30       &  8 &  31 &  39    \\
Sb,Sc        & 3,5      &  5 &   2 &   7    \\
Edge on disk & 50       & 11 &   7 &  18    \\
DIFRG        & 40       &  3 &  11 &  14    \\
Total        &          & 66 &  80 & 146    \\
\tableline
\end{tabular}
\end{center}
\end{table}

\subsubsection{Morphological classification scheme}

There are 146 red galaxies with redshifts, defined by
$U-B > 0.10$.  A first look at the sample suggested that a 
small number of categories could
accommodate them.  We then classified the red galaxies
into the categories listed in Table \ref{table-redmorph}.
Nearly all fell into one of three main categories:
spheroid-dominated, edge-on disk, and diffuse.
These categories were influenced by the appearance of typical
spheroid-dominated red-sequence galaxies at $z=0$ as well as the need
to identify other classes of objects whose redness may have
a different cause.  Examples of our classes are shown in 
Figure \ref{fig-redimages}.  

The spheroid-dominated category (hereafter called SRG,
for ``spheroidal red galaxy") contains objects from E to Sa.  
Such objects are presumably red because they are dominated by old stars.
Types were assigned to these galaxies based on axis ratio and 
visual degree of diskiness.  Spheroid-dominated objects 
that are too small to show structural details were assigned to 
a catch-all class of ``small spheroidals."  The small spheroidals
are more common at higher redshift, as expected from observational
effects.  Very few red galaxies
look like spirals of type Sb or later; only 7 objects
were identified as Sb or Sc.  The Sb's and Sc's are not included in the 
SRG group.

The edge-on class contains disky galaxies of all types that are
highly inclined; their inclination prevents a determination of
specific Hubble type.  These objects are presumably
reddened by dust, and several in fact show a dust lane.  Such
galaxies are found on the red sequence locally (see below),
and their occurrence at high redshift is not remarkable.

The third category, called DIFRG for ``diffuse red galaxy," is the most
interesting class.  We adopted this term when it became apparent that a
significant fraction of ``red smudgy''
objects lacked all trace of a spheroid, were
not edge-on disks, and yet were still red\footnote{
We use DIFRG rather than DRG to avoid confusion with the 
``distant red galaxies'' of van Dokkum \etal\ (2004).}.
Despite their redness,
morphologically these objects resemble late-type spirals or Magellanic
Irr I irregulars, though their absolute magnitudes are of course much
brighter, and their colors differ from 
low-redshift irregulars, which are generally very blue.  
Their central surface brightnesses are low compared to spheroids,
and their light distributions are much less concentrated, hence the name
``diffuse.''  The difference between these objects and SRGs is
striking, as shown in Figure \ref{fig-redimages}.  
The division between SRGs and DIFRGs is usually unambiguous, while
edge-on galaxies are harder to categorize as concentrated or
not, because of their orientation.  A number of authors have also
previously classified red galaxies into categories
roughly corresponding to early and late-type, usually in the
context of high-redshift extremely red objects (EROs) 
(Moriondo \etal\ 2000; 
%Stiavelli \& Treu 2000; 
Cimatti \etal\ 2003; Yan \& Thompson 2003; Moustakas \etal\ 2004).

We found that galaxies could be classified into 
spheroids and non-spheroids, orthogonal to whether they
were interacting or disturbed.
We therefore defined a separate classification parameter
largely based on asymmetry: in addition to
the SRG/disk/DIFRG categories above, each galaxy was also classified
as ``normal," ``peculiar," or ``interacting."  Normal objects
are fairly axi-symmetric or possess normal-looking spiral arms.  
To be classed as interacting, an object must have a clearly discernible 
separate companion, and the isophotes of the main galaxy must also 
be disturbed; plausible superpositions are classed as normal even though
the companion may be very close.  Objects that are neither axi-symmetric
nor interacting are called peculiar and are probably mostly late-stage 
mergers.  Of the 146 objects, 123 were normal, 20 peculiar, and
only 3 were classified as interacting.
None of the red galaxies had an interaction so profound that no 
SRG/DIFRG type could be assigned, although a few such mergers
are found among the blue galaxies.

Morphology and asymmetry types are presented
for all red-population galaxies in Table \ref{table-catalog}
using numerical codes
summarized in the endnotes and in Table \ref{table-redmorph}.  
Final classifications were made by BJW and CNAW,
and independently by SMF; good agreement was found.  Our types
also agree well those of Im et al. (2001).  Morphological
classification over a wide range of redshifts can be affected by
the redshifting of the bandpass (``morphological K-correction''), 
since galaxies look clumpier at bluer wavelengths due to 
starforming regions (e.g., Brinchmann et al. 1998).  However, our 
main goal is to sort spheroids from non-spheroids, not to make a
detailed classification of
later Hubble types, where blue stars dominate.  Since
HST $I_{814}$ remains fairly close to rest-frame $B$ out to $z \sim 1$, 
the K-correction relative to local spheroid Hubble types should be small.
This is consistent with our finding that the distinction between
SRG and DIFRG types is usually unambiguous and not dependent on
whether $V$ or $I$ images are used. 

\subsubsection{Color and redshift distributions of the galaxy classes}

The numbers of red galaxies found in the various morpholgical 
categories are summarized in Table \ref{table-redmorph}.
``Classic'' spheroid-dominated types are E, E/S0 through Sa, and
small SRGs.  Summing these, we find that $\sim 73\%$ of red galaxies
have bulge-dominated structure.  Roughly 12\% are edge-on disks,
presumably dust-reddened, and 10\% are DIFRGs, with a handful
of intermediate-type Sb's.  The fraction which are spheroid-dominated
does not differ significantly between low and higher redshift
when the sample is split at $z=0.7$.
However, the edge-on disks tend to be found at lower redshift
while the DIFRGs are mostly at higher redshift; both effects
are visible in Figure \ref{fig-colormag}.

We argue below, from analyzing Hubble types of local galaxies
in the RC3 catalog, that the DIFRG class is an essentially new kind
of object that does not appear in the local red population,
but is found there at high 
redshift.  It is therefore of interest to locate these objects 
in the CM diagram and chart their emergence versus redshift.
Figure \ref{fig-colormag} shows color and symbol 
codes for the various morphological
types of red galaxies.  The major color codes are
red for SRGs, green for edge-on disks and Sb's, and blue for DIFRGs.

The first conclusion is that SRGs and edge-on/Sb's are well mixed
in color at all redshifts, 
but there is a hint that DIFRGs may tend to be somewhat bluer, closer to
the valley dividing the red galaxies from blue.
This suggestion is confirmed by a study of valley galaxies just blueward
of the dividing line, which contain yet more DIFRGs.  The second
conclusion is that DIFRGs seem to be more prevalent at
high redshift and disappear with time: 
DIFRGs comprise 14\% of red galaxies beyond $z = 0.7$ but
only 4.5\% below $z = 0.7$.  Finally, the reverse effect is
seen for edge-on/Sb's, which comprise only 9\% of red
galaxies beyond $z = 0.7$ yet make up
17\% below $z = 0.7$.  This effect for edge-ons may be a selection
effect due to constraints from apparent size and surface brightness 
dimming, especially given their low surface brightnesses as shown
in Figure \ref{fig-colorsb}.  
Conclusions about redshift distribution of the types
are preliminary because the sample is small and because selection 
effects are not yet modeled.\footnote{We have checked the 
robustness of the morphological classes and their color and 
redshift distribution using the GOODS-N field with HST/ACS imaging,
which is deeper and has better spatial resolution.  We have 
classified a sample of $\sim 340$ red galaxies with redshifts,
to be reported in a future paper, and find the same 
general trends and proportions of spheroidal and diffuse red galaxies.
The most notable trend is that the better imaging reduces the 
fraction of ``small SRGs,'' allowing them to be classified as
either E,S0 or as bulges of Sa or Sb galaxies, with disks that
are only apparent in deep images.}

The percentage by number of classic early-type red galaxies in our 
sample is influenced by the intrinsic luminosities of the different
morphological types.  For example, if spheroidals are brighter
than diffuse galaxies, their contribution by number in
a magnitude-limited data set is magnified above the true ratio
of number densities.  However, their contribution to
the true luminosity density is also higher than the ratio of 
number densities, and these effects offset. 
A simple $1/V_{max}$
estimation of the luminosity functions of the various red
morphological classes yields an early-type E/S0/Sa/SRG percentage 
by number that is in general agreement with
the E/S0 fraction derived by Im \etal\ (2001) for GSS
galaxies down to $I=22$.  It also agrees with the GEMS team,
who used ACS images and photmetric redshifts
from COMBO-17 to determine that in the range $0.65<z<0.75$,
78-85\% of restframe V-band luminosity of red-population
galaxies is emitted by visually classified E/S0/Sa galaxies.
Further comparisons are beyond the scope of this paper,
given small numbers and the possibility of unknown
cosmic variance (our sample is small, and COMBO-17's narrow redshift 
range contains a large-scale-structure peak).

We also agree with COMBO-17 that many 
of the dust-reddened galaxies are edge-on disks, and
that the red population at $z=0.7$ is not dominated
by mergers or dust-enshrouded starbursts.  However, 
our DIFRG population, which is likely dusty, 
increases at redshifts beyond $z>0.7$.
Bell \etal\ (2003b) did not
note the existence of DIFRG galaxies as a separate class, but this 
could be due to their focus on the narrow redshift interval 0.65-0.75; 
by that epoch, most DIFRGs have disappeared from our sample.

The results found in these two surveys at $z \leq 1$
can be compared to a number of other studies which divide 
high-redshift red galaxies 
into classes by morphology or by spectral type.  These works
have generally selected extremely red objects (EROs)
based on, e.g., an $R-K>5$ color cut.  The samples contain mostly
intrinsically red galaxies at $z=1$ and higher, although
higher-redshift blue galaxies can also meet this color cut.

Moriondo \etal\ (2000) divided EROs into morphological classes
using HST images and concluded that 50-80\% were spheroidal,
but did not have redshifts or spectral information.
The K20 survey of EROs (Cimatti \etal\ 2002, 2003)
used spectroscopy and morphological classification from HST images
to divide EROs into three classes: E/S0, spirals, and irregulars, 
each about 1/3rd of the total of objects with identified redshifts.
The EROs with star formation were inferred to be dusty.
Yan \& Thompson (2003) classified EROs based on HST images and
found that 30\% were E/S0 and 64\% were disks, with 40\% of the
disks being edge-on and presumably dust-reddened.
Moustakas \etal\ (2004) classified morphologies of EROs in the
GOODS-S field with photometric redshifts
and found that 40\% were E/S0, 30\% later Hubble types, and
30\% irregulars; additionally, the fraction of normal Hubble types
was found to decline at higher redshift.

These studies classified EROs into early-type (spheroidal or
non-starforming) and non-early-type using various combinations
of visual morphology (spheroid/disk/other), Sersic index or 
concentration, and line emission.  We show here and below that 
the division of red galaxies into spheroidal versus DIFRGs and
edge-on disks is correlated with all these properties, and
surface brightness as well.  Early studies of ERO morphology
focused on the question of separating old spheroidal galaxies
from dusty mergers and massive starbursts, but as Yan \& Thompson
(2003) point out, ERO morphology is more varied than this
simple division allows.  Our survey shows a similar variety among
red galaxies at $z \sim 1$ (see also Abraham \etal\ 2004).
Our disks and DIFRGs may correspond to the disky
and late-type or ``irregular'' EROs; future studies which overlap
in redshift range and add $K$-band photometry will test this.

The fraction of red galaxies which are old and spheroidal 
has been a contentious issue in studies of EROs.  Reconciling
surveys with different selection and morphological criteria
is beyond the scope of this paper, but we can make
relative statements about the redshift trends in the 
red galaxy population.  Moustakas \etal (2004) classified
ERO galaxies in the GOODS-S field with photometric redshifts,
a sample concentrated in the range $z\sim1$ to 2.  
Near $z=1$, that sample
contains mostly normal Hubble types, divided roughly 50-50
between early types with no visible disks and late types
with disks.  This is consistent with our results because
both we and Bell \etal\ (2003b) group Sa
galaxies with early types; these (and S0's)
show disks and would be classed as late-type
by Moustakas \etal\  By $z=1.5$, however, the number
of normal Hubble types in GOODS-S has declined, and the sample is
dominated by irregular and ``other'' types, which, as
the names imply, do not fit well onto the normal Hubble
sequence.  From inspection of their Figure 2, some of these
could be DIFRGs.  The DIFRGs we find may be the $z<1.5$ equivalent
of dusty EROs; if so, they provide an excellent laboratory
to study dusty EROs at redshifts where spectral
features are accessible in the optical.

In sum, the number of normal, early Hubble types among
red galaxies appears to grow steadily with the passage of time toward
$z\sim 0.5$.  RC3 data discussed below appear to confirm
a further growth from that epoch to now. 

The finding here and in Bell \etal\ (2003b)
that red galaxies are mostly spheroid-dominated 
out to $z \sim 1$ prompts us to check, by analogy with
local E/S0 galaxies (e.g., Small \etal\ 1999, 
Shepherd \etal\ 2001), whether they are also more
highly clustered and preferentially
populate dense regions.  A quick check comes from
the pie diagram, Figure \ref{fig-piediag}, 
where red galaxies are shown as red
dots.  The visual impression is that red
galaxies do indeed inhabit higher density environments, 
an impression that is quantitatively confirmed by
data from the DEEP2 survey (Coil \etal\ 2004; Cooper \etal\ in preparation).

Blue galaxies, those with $U-B<0.1$, show a 
variety of morphologies.  Many blue galaxies are more diffuse
and have lower surface brightness than red galaxies.  There are
many disks and very few galaxies that look like spheroid-dominated,
E/S0 types.
A substantial number are clumpy or peculiar; since star formation
tends to be higher in blue galaxies, the morphological
K-correction may play a role, but since the $I$-band
image corresponds nearly to restframe $B$ at our outer limit
$z \simeq 1$, the bias compared to local Hubble types should
be small (van den Bergh \etal\ 2000).  A high fraction
of blue peculiar galaxies
has been noted at high $z$ by 
many authors (e.g., Brinchmann \etal\ 1998)
and may have persisted to as recently
as $z = 0.3$ (see review by Abraham \& van den Bergh 2001).  

\subsection{Concentration versus color}

\begin{figure}[ht]
\includegraphics[width=3.5truein]{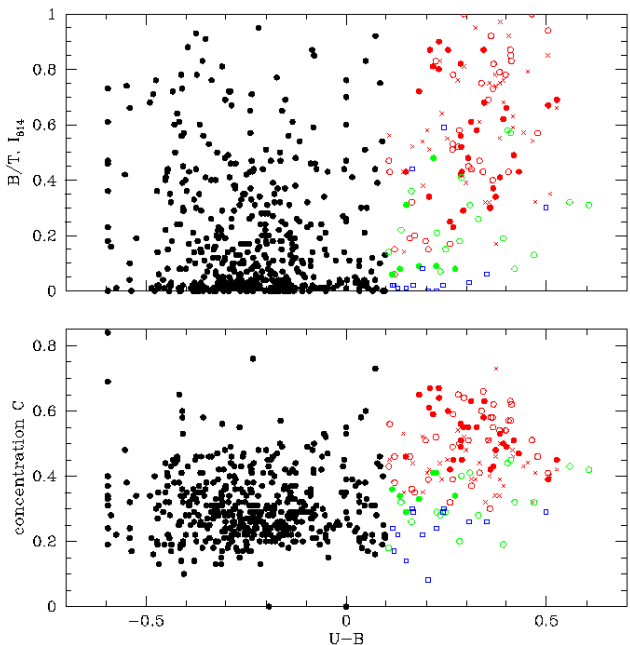}
\caption{
Bulge/total ratio and concentration parameter $C$
measured in the $I_{814}$ images
versus restframe $U-B$ color.  Both quantitative
morphology parameters confirm the visual assessment
that red galaxies tend to be bulge-dominated
with high central concentrations, while blue galaxies are
more disky with lower concentrations.  
Red galaxies are coded for morphological type as
in Figure \ref{fig-colormag}: red = spheroidals; green = disks;
blue = DIFRGs.
Within the red population, morphological type is further correlated with
concentration: low concentration and low B/T galaxies are disks
or diffuse DIFRGs; high B/T galaxies are
SRGs.
}
\label{fig-colorconcen}
\end{figure}

Quantitative structural parameters that correlate well with
visual morphology for local galaxies are bulge/total
ratio and central concentration.
$B/T$ is measured by fitting a model bulge and disk
to the light distribution, while several definitions of
concentration exist.
Figure \ref{fig-colorconcen} plots two measures
computed by {\tt gim2d} (Simard \etal\ (2002):
the bulge/total ratio, and
the $C$ concentration index defined by Abraham \etal\ (1994),
versus $U-B$ color.  The $C$ index is effectively a ratio
of the flux inside a given fraction of a fiducial total radius 
to the flux inside that total radius, and is
about 0.25 for a pure exponential disk at the typical resolution
in this sample.
%\comment{Check this for c0.1 versus c0.3; or use c(0.3);
% dependent on resolution??}

It is clear that color and concentration are correlated: 
galaxies form two distinct groups in both
quantities, showing that color 
bimodality also mirrors quantitative structure.
Blue galaxies have low
concentrations and are largely clustered around pure exponential disks,
with a tail to more concentrated galaxies.  Red galaxies have
high concentrations, and show a range in $B/T$ above $\sim 0.2$.  
Even in the red galaxies, few objects are found with a
pure bulge or pure $r^{1/4}$ profile.
This correlates with the findings of Im \etal\ (2001),
using GSS galaxies down to $I=22$,
that morphologically normal
pure E's and E/S0's actually exhibit a wide
range in $B/T$.  The threshold in that paper was set
at $B/T>0.4$, which recovers most E and E/S0 galaxies based
on the data here.  Our results also agree with
Bell \etal\ (2003b), who used the Sersic $n$ index
to measure the concentration of distant GEMS galaxies, obtaining the
same conclusions as from visual classifications. 

For red galaxies, our morphological
typing is confirmed by quantitative structural
measurements.  Sb and edge-on
disks (green points) have lower concentrations on average
than spheroidal types (red points).
DIFRGs (blue points) have yet lower concentration; in fact the
DIFRGs are clearly unusual, since they violate the relation
between concentration and color seen from blue galaxies through 
transition objects to red SRGs.
We also examined the relation of red/blue color and
morphological type to asymmetry index, but found no clear
correlation (see also Wu 1999; Wu, Faber \& Lauer 2003).

The color-concentration relationship for distant
galaxies in Figure \ref{fig-colorconcen}
is very similar to that found for galaxies
at low redshift. As an indicator
of concentration, Blanton \etal\ (2003)  
used a single-component Sersic fit to 
the photometric profiles of SDSS galaxies (Sersic $n=1$ 
for exponential profiles and 4 for de Vaucouleurs profiles).  
Blue SDSS galaxies are strongly clustered around $n=1$, 
while red galaxies are more concentrated and range
from $n=2.5$ to 4.  Bell \etal\ (2003b)
studied SDSS galaxies using the inverse Petrosian ratio $C_r$, 
defined as $r(90)/r(50)$, and found similar results.
From both near and distant studies, then, it
appears that high central concentration
is a prerequisite for residency on the red sequence,
at least for galaxies that are red because
of old stellar populations rather than dust.  

Figure \ref{fig-colorconcen} shows that a 
highly concentrated light
component is also present in 
some blue galaxies.  For fitting purposes, that component
is called a ``photo-bulge'' in the terminology of
Simard \etal (2002), but it is not necessarily a true bulge
in the sense of an old, red spheroidal stellar population.
Koo \etal\ (2004) have examined
GSS galaxies individually to identify morphologically
normal bulges.  Almost without exception, true bulges
are found to be red compared to the surrounding galaxy,
while genuine blue bulges are rare.  Im \etal (2001) 
found that blue centers live in galaxies with higher
asymmetry, and that these blue-centered galaxies have small
masses and central bursts of star formation.  We discuss
the colors of ``bulges'' in blue galaxies further in 
Section \ref{sec-bulgedisk}.

\subsection{Color and surface brightness}

A stellar population fades significantly as it reddens with age.
B03 showed that because the most luminous red galaxies are 
brighter than the blue galaxies, the red galaxy population
found in their survey cannot be produced by fading of the 
observed blue galaxies.  Our color-magnitude distribution is
similar and supports this result, as does the color-stellar mass
distribution of Figure \ref{fig-colormass}.  Given the
difference in structure between red and blue galaxies,
it is also useful to look at the relation between color
and surface brightness.

\begin{figure}[ht]
\includegraphics[width=3.5truein]{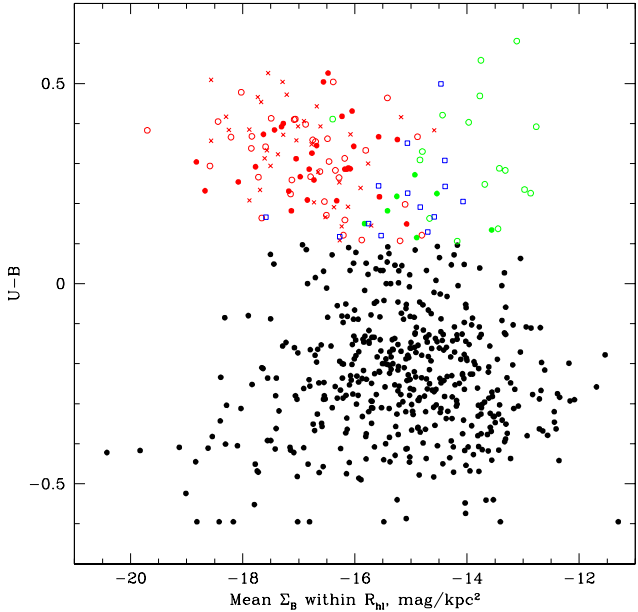}
\caption{Mean surface brightness vs. total $U-B$ color.
$\Sigma_B$ is the mean B surface brightness within $R_{hl}$
in magnitudes per kpc$^2$.  Points are coded for morphological
type as in Figure \ref{fig-colormag}.  Surface brightness is
strongly correlated with red galaxy morphological type:
spheroidals (red) have highest $\Sigma_B$, DIFRGs (blue) are
lower, and edge-on disks (green) are lower still.}
\label{fig-colorsb}
\end{figure}

We define a measure of mean surface brightness 
within the half-light radius $R_{hl}$ in kpc,
$\Sigma_B = M_B + 2.5~{\rm log}~2 + 2.5~{\rm log}~\pi R_{hl}^2$, 
which is applicable
to all galaxies independent of bulge/disk ratio.
Figure \ref{fig-colorsb} shows the relation between
galaxy color and surface brightness.  
Red and blue galaxies have rather different distributions
in surface brightness, and in the red galaxies, surface 
brightness is strongly correlated with morphological class.

Among the blue galaxies, there is a tail of high surface 
brightness objects with mean $\Sigma_B<-17$.  These tend
to be small, compact galaxies, similar to those described by
Phillips \etal\ (1997).
Aside from these compact galaxies, red spheroidal galaxies
have a typical mean surface brightness that is
higher than blue galaxies by $\sim 1.5$ magnitudes.
This division is similar to the correlation between color and 
concentration and is clearly associated.  
When converted to {\it mass} surface density, the difference
is even higher.  

Within the red
population, spheroidal red galaxies (red points) have much higher
surface brightness on average than disky galaxies (green
points) and diffuse red galaxies (blue points), by $\sim 2$
magnitudes.  The division in surface brightness is quite striking --
of course, surface brightness was one criterion used in the
morphological classification -- and suggests that surface brightness
might be useful as a proxy for morphological type among red
galaxies.  The edge-on disks are quite low in surface 
brightness and this is probably a major reason that they are more 
difficult to find at higher redshifts, since surface brightness
selection can have a strong effect on redshift distribution
of a sample (Simard \etal 1999).

The surface brightness division is further proof that 
the observed blue galaxies cannot 
redden and fade into the observed red galaxies.  The predecessors of 
high surface brightness red galaxies would have to be large objects
with much higher SB than nearly all blue objects that we see.  

\subsection{Color-magnitude diagrams of bulges versus disks}
\label{sec-bulgedisk}

\begin{figure*}[ht]
\centerline{
\includegraphics[width=5.5truein]{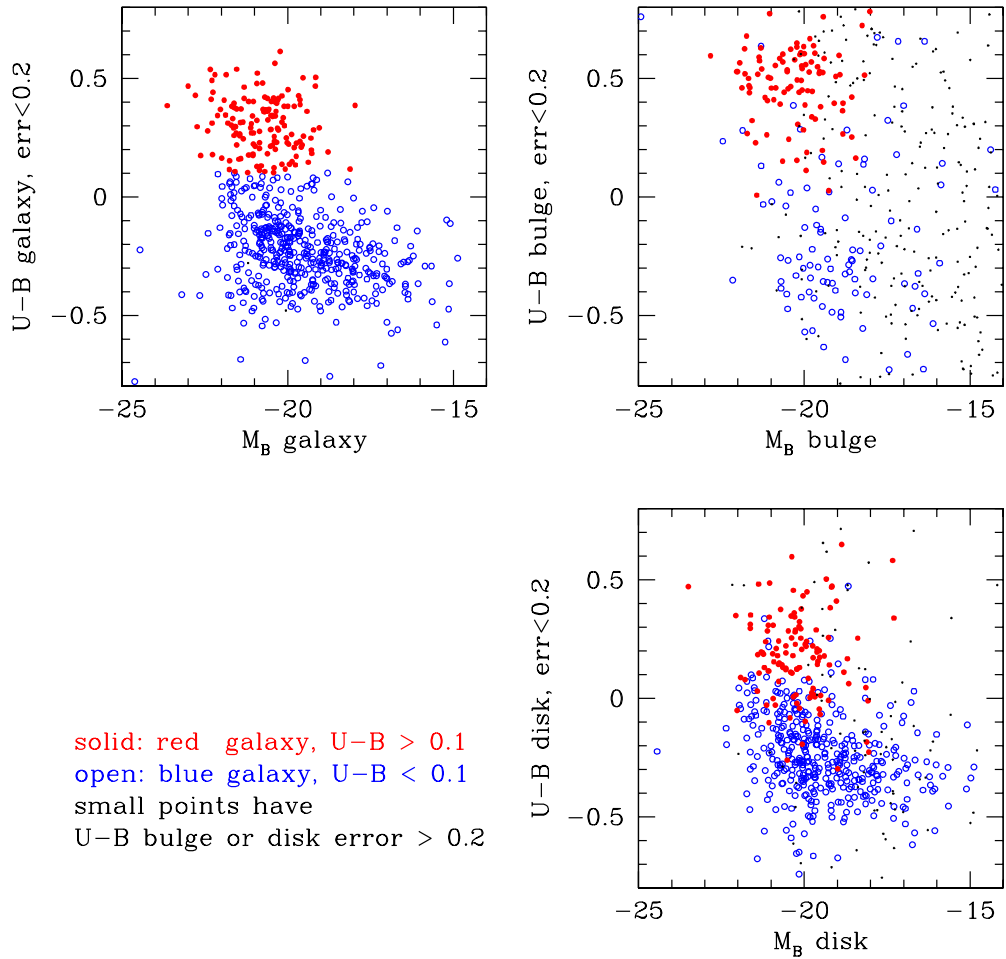}
}
\caption{Color-magnitude diagrams separately for bulges,
disks, and total galaxies.  Points are coded by
integrated total galaxy color: red galaxies are red filled circles,
blue galaxies are blue open circles.  Red and blue galaxies
behave oppositely in this figure.  Red-population
galaxies show an internal color gradient in the sense
that bulges are {\it redder} than disks (i.e.,
the inner parts are redder than the outer parts),
consistent with the run of color in local
early-type galaxies.  Blue galaxies show
less change between disk and bulge, but, when bulges 
are well measured, they appear 
to be systematically {\it bluer} than surrounding disks.
This is consistent with the conclusion
that the ``bulges'' in blue galaxies 
are in fact central star-forming regions.}
\label{fig-buldiskcolor}
\end{figure*}

The photometric decompositions by
Simard \etal\ (2002) make it possible to plot color-magnitude
diagrams for bulges and disks separately.
These are shown in Figure \ref{fig-buldiskcolor}, where
points are color-coded by the
integrated color of the whole galaxy.  For many 
objects with low $B/T$, the bulge color is not well measured,
and these are shown as small points.  

For galaxies with reliable 
bulge/disk decompositions, there is a distinct difference 
in behavior between red and blue galaxies.  
Red galaxies have luminous red bulges, with
mean bulge color $U-B=0.5$, significantly redder than the
mean total color of 0.35, and much
redder than the mean disk color of 0.20. 
These are bulges in the traditional sense: spheroids
with old stellar populations.
Disk components bluer than the bulges imply
a radial color gradient in the sense that the outer parts
are bluer, in agreement with local early-type galaxies.
The physical origin of these color gradients could be
that the outer parts are younger, more metal-poor,
or both.  

The behavior of blue galaxies is opposite.
Most of the ``bulges''
in these systems are quite blue, with mean color $U-B=-0.3$,
which corresponds to vigorous star formation. 
Furthermore, Figure \ref{fig-buldiskcolor} 
shows that blue central components are
systematically {\it bluer} than their outer disks.
Both facts support the notion
that blue galaxies with blue ``bulges'' are actually
hosting central regions that are actively forming young stars.
For example, the distant compact narrow emission-line galaxies
of Guzman \etal\ (1998) tend to have a high bulge fraction,
but colors and $M/L$ indicating a central starburst.

A final fact, noted by Koo \etal\ (2004)
and visible in Figure \ref{fig-buldiskcolor}, is
that even the brightest blue ``bulges'' are actually
about 1 magnitude dimmer than red bulges.  
Since a stellar population both fades and
reddens as it ages, these blue centers {\it cannot} be the
progenitors of the observed red $r^{1/4}$ bulges;
they are too faint.  If after fading, the blue center will
deviate from local relations for luminous early-type bulges, then
it cannot be a proto-luminous {\it bulge}, though it could perhaps
produce a fainter or exponential bulge as found in nearby late type
galaxies.  Because we do not find 
any sufficiently bright concentrated blue sources, and few of
intermediate color, we conclude that
either the observed red bulges formed well before $z=1$
(see also Abraham \etal\ 1999), or
that massive bulge formation involves growth
of mass by mergers after $z = 1$, or that 
massive bulge formation below $z=1$ is happening in sites of highly 
obscured star formation.
These conclusions for red bulges parallel the conclusions on the
formation of entire gaalxies in the red population reached by B03.

\subsection{Line emission versus type and color}
\label{sec-emission}

\begin{figure*}[ht]
\centerline{
\includegraphics[angle=-90,width=6.5truein]{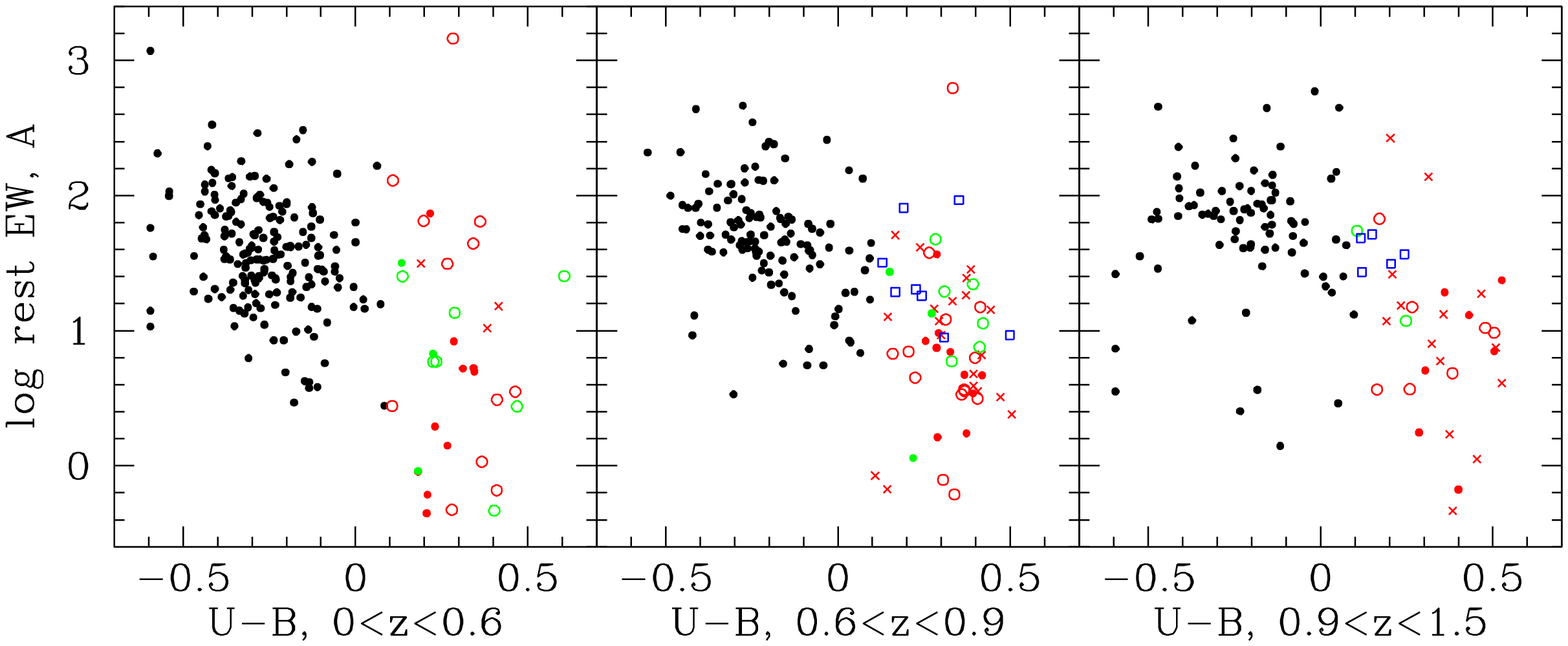}
}
\caption{Log equivalent width in \AA\ of the strongest observed emission 
line versus $U-B$ restframe color, in three redshift bins.  
EW is strongly related to color for
blue galaxies, perhaps less so for red.  Red galaxies are coded for
morphological type as in Figure \ref{fig-colormag}: red = spheroidals;
green = disks; blue = DIFRGs.
Diffuse red galaxies (blue points) have generally higher emission
than spheroidal red galaxies, suggesting that DIFRGs are 
dust-reddened rather than old.}
\label{fig-colorew}
\end{figure*}

Most evidence up to this point indicates that, to first
order, the distant red and blue populations are composed of
galaxies that broadly resemble 
their low-$z$ counterparts.  We might therefore also expect to find
a strong trend between restframe color and star-formation rate,
as seen in local galaxies.  This
is confirmed in Figure \ref{fig-colorew},
which shows the log restframe equivalent width (EW)
of each galaxy's strongest emission line plotted against galaxy 
color.  The strongest line is usually \ha\ 6563 at very low $z$, 
\oiii\ 5007 at $z \sim 0.5$, and \oii\ 3727 at $z>0.6$.
Strength has been defined here as highest intensity 
in DN units, not in EW, although the two are closely related.  
The median error in restframe EW is 6.2 \AA, and the median
error on log EW is 0.08 dex.  
%A few objects with poorly measured EW
%due to low continuum levels are omitted.

Figure \ref{fig-colorew} shows the expected decline in EW at 
redder color for the sample as a whole, and an approximately
linear relation between log EW and color, with substantial scatter.   
We can further ask if there 
is any trend between EW and morphology for the red galaxies.
In Table \ref{table-ewmorph} we show the median EW by
morphological type for galaxies with well measured EWs.

\begin{table}[t]
\begin{center}
\caption{Median emission EW for galaxies by morphological class}
\label{table-ewmorph}
\begin{tabular}{lrrr}
\tableline\tableline
Type               &   Number   & median  & median  \\
                   &            &  $z$    &  EW, \AA \\
\tableline

SRGs (E/S0/Sa/small)  &  100    &  0.809  &  6.6 \\
SRGs, $z>0.7$         &   59    &  0.878  &  7.5 \\
Sb/edge-on            &   21    &  0.604  & 11.8 \\
DIFRGs                &   13    &  0.814  & 31.3 \\
Blue, $U-B<0.1$       &  411    &  0.632  & 45.1 \\
\tableline
\end{tabular}
%\tablenotetext{a}{Only galaxies with well measured EW are included.}
\end{center}
\end{table}

Because the diffuse red galaxies
are found only at $z>0.64$, we also tabulate the EW for spheroidal 
red galaxies with $z>0.7$ to make a fairer comparison.  At $z>0.7$, the 
strongest-line EWs all refer to [O II] 3727.  The median EW
of diffuse red galaxies is four times higher than that of
spheroidal red galaxies, confirming the impression seen in Figure 
\ref{fig-colorew}.  The DIFRGs are bluer than the SRGs, so can
be expected to have higher emission given the color-EW relation,
but Figure \ref{fig-colorew} suggests that the DIFRGs also lie 
at higher EW than the mean relation.
The median EW of DIFRGs is high enough 
to indicate reasonably active star formation.  We conclude that
DIFRGs are not dominated by old stellar populations,
and are reddened mainly by dust.  However, since they do not have
the characteristic appearance of the edge-on disks,
they are a distinct, new class.

\subsection{Scatter in color on the red sequence}
\label{sec-scatter}

We have seen that the red population contains a fair
range of morphologies.  Most objects are morphologically 
similar to nearby early types from E to Sa, but 
dust-reddened, irregular, and interacting objects can 
also fall on the red side of the color bimodality.
This range of morphology could produce 
scatter in other properties, such as color.
Red-sequence cluster galaxies generally fall on a tight
color-magnitude relation (CMR), both locally (Bower \etal\ 1992) 
and at high redshift (van Dokkum \etal\ 2000).
It is clear from Figure \ref{fig-colormag} that our sample 
of red field galaxies does not obey a perfectly narrow
color-magnitude relation, nor does it have an obvious slope.  
However, measuring the slope and scatter are
not easy because the distribution of galaxies about
the red sequence is asymmetric.  
On the blue side of the peak,
the distribution of points blends into
the red wing of blue galaxies, and the presence
of these objects will increase the apparent scatter
of the red sequence per se unless these objects
are allowed for or the sample is truncated.  

B03 dealt with this by assuming a fixed CMR slope,
truncating the sample to $U-V_{AB}>1.0$, and
calculating a robust biweight
estimator (Beers \etal\ 1990) of the evolving CMR zeropoint and
the scatter on the red sequence.
They found a Gaussian scatter on the red sequence of 0.15 mag
in $U-V$ rest, or 0.09 mag transformed to
$U-B$ rest, which they attributed largely to 
photometric redshift error.  If the CMR zeropoint evolution
were not subtracted, the scatter would increase only marginally
to $\sim 0.10$ mag in $U-B$.
We carried out a similar calculation
by truncating our sample to $U-B>0.1$, the trough of
the color distribution, and computing
the biweight scatter.  We find a scatter of 0.116 mag.
This number changes only slightly to 0.112 mag when we exclude the $z<0.4$ 
objects with poorer K-correction, and does not change if we subtract 
a CMR slope in $U-B$ of 0.04 (van Dokkum \etal 2000).  The scatter 
in $U-B$ for only spheroidal morphological types is nearly as
large, 0.108 mag, and is reduced only to 0.100 mag when $z<0.4$ 
objects are excluded; subtracting a CMR slope again
does not change the scatter.

Our CMR scatter is therefore comparable to that of B03.  While our
K-corrections are less accurate since we have fewer filters,
we eliminate the errors due to scatter in photometric redshifts.
The derived scatter is large compared to the known sources
of error, which were enumerated in Section \ref{sec-colors}.
Median errors of measurement in $V-I$ are only
0.055 mag, which converts to 0.03-0.04 mag in $U-B$ for
a typical galaxy in the range $z = 0.4$-1.0.
Correcting the raw scatter of 0.112 mag for this
observational error yields a net true scatter of
0.105 mag.

The other potential source of error is the K-correction in
converting apparent $V-I$ to restframe $U-B$.
The errors inherent in K-correction are
small in the redshift interval $z = 0.4$-1.0 because
$U$ and $B$ transform almost directly 
into $V$ and $I$ at $z=0.8$.  We have checked this by
identifying groups of red galaxies within
individual redshift peaks.  Within a peak,
comparing the behavior of the {\it apparent} color-magnitude relation
in $V-I$ vs. $I$ to the C-M relation in $U-B$ vs. $B$
separates the scatter in the measured apparent colors from the
scatter introduced by the K-correction procedure. 

This test shows that our scatter in the red sequence colors has two
causes.  In the redshift peaks at $z=0.56$, 0.65, 0.81, 0.91, and 0.99
the apparent $V-I$ colors have a scatter of $\sim0.1 - 0.2$ mag.
The scatter in $V-I$ increases with redshift, in part due to the
changing transformation of rest to apparent colors; at 
$z\sim 0.9$, the slope of $U-B/V-I$ is 0.6.
Here the K-correction procedure works well, and the scatter in $U-B$
is clearly due to input scatter in $V-I$.
In the redshift peak around $z=0.28$, the 
red galaxies form a tighter relation in $V-I$ vs. $I$,
but the K-correction at this low redshift causes 
$U-B$ to be a stronger function of $V-I$ with slope 1.0,
amplifying photometric errors and the scatter in $U-B$.  
%This merely reflects the
%fact that the slopes of galaxy SEDs vary more in the UV-blue
%than in the visual bands.
The upshot is that, beyond $z=0.4$, the scatter is real and 
reflects real scatter in input $V-I$.

The salient conclusion is that the scatter in the color of
galaxies in the red population, 0.10 mag in $U-B$ in the range
$z = 0.4$-1.0, is intrinsic and not
due to observational error.  Our scatter of 0.10 mag is close 
to the 0.09 ($U-B$) mag found by B03 based on their extensive
17-filter ground-based photometry.
While the scatter in the CMR of both local and high-$z$ {\it cluster}
galaxies is small, $\sim 0.03$-0.04 mag when converted to $U-B$ 
(Bower, Lucey, \& Ellis 1992; van Dokkum 
\etal\ 2000), the scatter in a local sample of {\it field} E/S0
galaxies is larger, $\sim 0.06$ mag (Schweizer \& Seitzer 1992).
The local scatter is increased further, to 0.10 mag, 
by including all Hubble types 
and not correcting for internal extinction, as shown using RC3 galaxies 
below in Section \ref{sec-localcomp}.  Notably, the scatter of
DEEP1 red galaxy $U-B$ color is significantly larger than the
scatter in DEEP1 red {\it bulge} color of only $\sim 0.03$ mag
(Koo \etal\ 2004).
We conclude that the red-sequence scatter we and B03 find is
comparable and reflects real scatter in red field galaxy colors,
both nearby and at high redshift.

\section{Comparison to local data}
\label{sec-localcomp}

\begin{figure*}[ht]
\centerline{
\includegraphics[angle=-90,width=6.0truein]{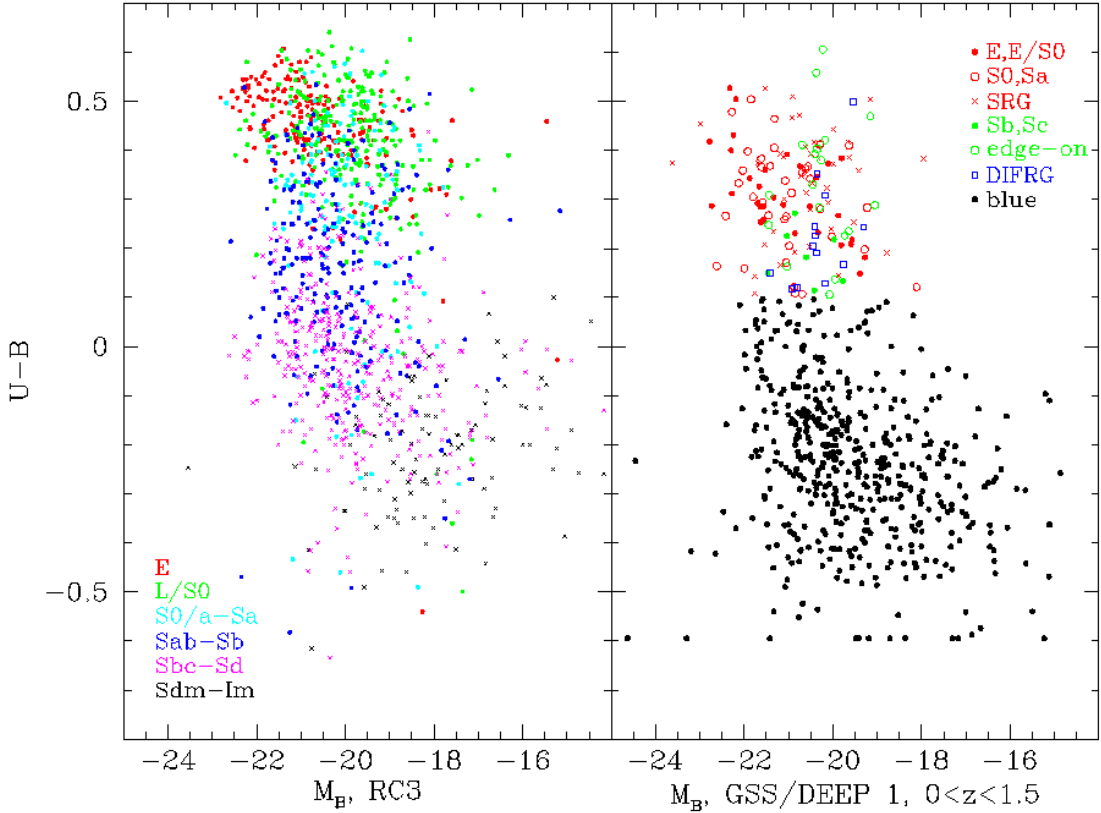}
}
\caption{Color-luminosity diagram for RC3 galaxies 
compared to DEEP1/GSS galaxies.  Panel (a): RC3 galaxies
with $U-B$ values, integrated $B$ magnitudes, redshifts,
and $B_T^0 < 13.5$.
%The methods used to calculate $U-B$ and $M_B$ 
%are described in the text.
%Color codes for RC3 morphological T-type: red = E; green = L; 
%cyan =  S0/a and Sa; blue =  Sab-Sbc; magenta = Sc-Im.
Color codes for RC3 morphological T-type: red = E; green = L; 
cyan =  S0/a and Sa; blue =  Sab-Sb; magenta = Sbc-Sd; black = Sdm-Im.
The local red seqence shows the well known slope versus magnitude.  
It is dominated by Hubble types E-Sa, similar to
the distant red population.
(b) The CM diagram for DEEP1/GSS galaxies to the same
scale. Red galaxies are coded by morphological type as in Figure
\ref{fig-colormag}.}
\label{fig-colormagrc3}
\end{figure*}

We conclude this discussion of distant galaxies 
by comparing to the color-magnitude
diagram of local galaxies as represented in the RC3.  
Figure \ref{fig-colormagrc3} plots
the CM diagram of galaxies in the RC3 (de Vaucouleurs
\etal\ 1991) with integrated $U-B$ measurements,
Galactic extinction $A_B<0.3$, and apparent magnitude
$B_T^0 < 13.5$, to which the RC3 is reasonably complete.
Absolute magnitudes are computed from 
$B_T$ and radial velocity in the CMB frame.  Galaxies
with $v_{CMB}<0$ are excluded.  We plot 
total magnitude $M_B$ and color $(U-B)_T$, after correcting
for Galactic extinction but not for internal extinction, in order
to be comparable to the distant galaxy sample.

Use of the RC3 data as a local sample has two advantages.
First, if we have correctly reproduced the $UB$ system for
distant galaxies, no further color or aperture
corrections are needed, as the RC3 galaxies are on
that system. 
Second, most RC3 galaxies have Hubble types, which
allows us to characterize the morphologies of local red-sequence 
galaxies in the
same terms that we have set up for distant red
galaxies in Section \ref{sec-morphology}.

Figures \ref{fig-ubhist} and  \ref{fig-colormagrc3}(a) 
demonstrate the same sort of color bimodality identified in
local SDSS galaxies (e.g., Strateva \etal\ 2001, 
Hogg \etal\ 2003, Kauffmann \etal\ 2003).  The red population peaks around
$U-B=0.45$ and shows the well known 
early-type color-magnitude relation.  The red peak is 
clearly divided from the blue galaxies, which peak separately
around $U-B=-0.05$.  No comparable color-magnitude relation 
is visible for red galaxies in
our DEEP1/GSS data, but B03
show one clearly out to $z\sim1$ (with local slope).
Our sample may be too small to show a relation given
the large scatter.

The right-hand side of Figure \ref{fig-colormagrc3}
compares the CM diagram of GSS galaxies
to the same scale.  Plotting them side-by-side 
highlights the color evolution that has occurred in galaxies since
$z \sim 0.7$, which is also shown in Figure
\ref{fig-ubhist}.  
The peak of the red population has shifted redward with time 
by about 0.11 mag in $U-B$, the
valley has moved by the same amount, and the blue peak has shifted by about
0.24 mag.  Selection effects cause faint galaxies to be
missing from both distant and local samples, although
the effects act in the same direction in both samples.  
Comparing the distribution of 
points and the color histograms of Figure \ref{fig-ubhist} suggests
that the RC3 sample of blue galaxies is more heavily weighted
toward brighter (thus redder) galaxies than the DEEP sample, which 
exaggerates the evolution.  The color evolution of blue galaxies 
can be measured more reliably with better local samples, and with larger
distant samples which measure evolution internal to the samples,
from COMBO-17 and from DEEP2 (Willmer \etal\ 2004, in preparation).

Our color evolution of 0.11 mag in $U-B$ for the red population,
from its median redshift of 0.8 to now, can be compared
to the color evolution of red sequence galaxies measured by B03.  
Their Figure 2 shows an evolution from $z=0.8$ to 0.25 of 0.26 
mag in $U-V$, and extrapolates to 0.35 mag to $z=0$,
corresponding to 0.16 and 0.22 mag in $U-B$ respectively.  Both
numbers are larger than what we find.  However, Figure 2 of B03
also shows that their red galaxies at $z=0.3$ are {\it already redder}
than the colors of local galaxies from SDSS.  In contrast, the color
evolution from the COMBO-17 red galaxies at $z=0.8$ to SDSS
galaxies at $z=0.05$ is only 0.15 mag in $U-V$ directly, or 0.1 mag 
converted to $U-B$,
which agrees with our result.  B03 noted that aperture effects
may have caused their low-redshift galaxy colors to be too red,
which could explain the difference.  The color evolution
that we find is consistent with the small color evolution found by
Gebhardt \etal\ (2003) and by Koo \etal\ (2004), and is
only half that predicted by passive evolution
models with $z_{form} \sim 3$.  The amount of color
evolution in red galaxies is critical in determining their
stellar population histories and the amount of mass buildup
on the red sequence.  These questions will be addressed with
larger samples and better restframe colors in future DEEP2 papers.

Figure \ref{fig-colormagrc3}(a) shows Hubble types for RC3 galaxies.
The RC3 red sequence is composed mainly of E and
S0 galaxies, while Sa
galaxies are centered on the valley between the peaks.
The local population redder than the valley minimum therefore
contains mostly Hubble types E to S0, with a smattering of Sa's.
This resembles the makeup of
the distant red population, which consists mainly
(75\%) of spheroid-dominated SRGs
(Section \ref{sec-morphology}, also Bell \etal\ 2003b).

However, roughly 25\% of our distant red galaxies are {\it not} typed as
SRGs.  Of these, nearly all are either edge-on disks (some of which may
be SRGs) or are DIFRGs.  We have examined the RC3 sample
to see if such objects are present on the local red sequence as well.  
A small contaminating population
of intermediate Hubble types (Sab to Sc) is indeed found, at the few
percent level, and most of these interlopers are edge-on or 
close to it and are likely reddened by dust. 
%such types are usually too blue to lie on the red sequence.  
These are presumably the
analogs of our distant edge-on galaxies.  The number of these
edge-on middle Hubble types is about the same in both the DEEP1 and RC3
samples as a fraction of the blue galaxy population.

The DIFRGs are another matter.  We have seen that they 
tend to populate the valley
between the red and blue population, the region that locally
is occupied by Sa's.  However, DIFRGs structurally
do not resemble Sa's, because they completely lack a large spheroid.
The most plausible local analogs to DIFRGs in structure are types
Sc and later.  But there are only three Sc's on the entire RC3 subsample
red sequence (out of 517 galaxies), and all three are 
edge on and dust-reddened.
DIFRGs, we have noted, are distinct from classic edge-on disks, because
they are not as flattened.
The most plausible local red-sequence
analog to DIFRGs is perhaps the dusty, irregular starburst
galaxy M82, which has $U-B=0.28$ and whose light is not strongly 
concentrated in the $B$-band (although it is more concentrated
in $K$ images).  M82 itself is highly inclined but is likely
to be reddened even if not viewed edge-on.  Other than M82, there
do not seem to be common local red-sequence analogs to distant 
DIFRGs, which is consistent with their apparent disappearance at 
low redshift in our sample, as noted in Section
\ref{sec-morphology}.  It {\it is} possible that DIFRG analogs could exist
locally at the {\it same color} as high-z DIFRGs, $U-B \sim 0.15$.  In
that case they would not be on the local red sequence, but would be
a minority within the red edge of the blue 
population, which is mostly disk galaxies with bulges in the RC3
sample.   Thus, both the nature of high-z DIFRGs and what they evolve 
into remain unclear.

The next point to examine is the color scatter of galaxies
about the RC3 red sequence.  Applying the same method used in Section 5.8 to
reject outliers, we find that the biweight scatter about the red peak is
0.10 mag in $U-B$ for all galaxies with $U-B>0.2$,
if no CM relation is subtracted, and
decreases insignificantly when a CM relation of slope 0.025 is subtracted
and the sample is cut by a line 0.25 mag blueward of the CM relation.
To parallel the treatment of the distant samples, all Hubble types are
included, and no corrections for internal reddening have been made.
This RC3 scatter is essentially the same as the 0.11 mag we find 
for the GSS galaxies
(with no CM relation subtracted), and the 0.09 mag found by B03 (with 
an evolving CM relation subtracted).  We conclude that an rms
scatter in $U-B$ of $\sim0.10$ mag typifies field red-sequence galaxies
at {\it all} redshifts below $z \sim 1$.  This is significantly larger 
than the scatter of 0.03-0.04 found for red, early-type cluster galaxies, 
both near and far (e.g., Bower \etal\ 1992; van Dokkum \etal\ 2000).

Finally, the reader may have noticed that the number of red galaxies
relative to blue galaxies is only about half as large in the GSS CM
diagram as that in the RC3.  One factor that discriminates against
distant red galaxies is the R-band selection limit, shown in Figure
\ref{fig-colormag}, which preferentially selects against red galaxies
at high redshift.  On the other hand, a real increase in the number of
red galaxies with time would be consistent with the increase in total
red {\it stellar mass} reported by B03 below $z = 1$.  This topic is
discussed further in Willmer \etal\ (2004, in preparation), where we 
derive luminosity functions from DEEP samples.

%\comment{We have not mentioned the fact that local galaxies peak
%at a color redder than almost all of the DEEP galaxies, but about
%equal to the DEEP red bulges.}

\section{Conclusions}

The DEEP1 survey of the Groth Strip provides redshifts and spectra for
over 600 galaxies out to $z\sim 1$, with HST/WFPC2 imaging in two
filters; the median redshift is $z=0.65$.  Large scale structure
walls are visible in the space distribution to $z\sim 1$.

The DEEP1 galaxy sample shows a bimodality in the color-magnitude diagram
similar to that found locally in
the SDSS (Blanton \etal\ 2003; Kauffmann \etal\ 2003)  and to $z \sim 1$ in
COMBO-17 (B03).  The brightest red galaxies are somewhat
more luminous than the brightest blue galaxies, so that the
observed blue galaxies cannot fade to produce the brightest 
red galaxies, as discussed by B03.  This difference between the most
luminous red and blue galaxies is even stronger in terms of
stellar mass, as shown in Figure \ref{fig-colormass}.
The processes of star formation, quenching, and mass assembly in red 
galaxies will be studied at greater length in future surveys, including 
DEEP2 measurements of the
luminosity function (Willmer \etal\ 2004, in preparation). 

The present sample does
not permit a measurement of the red galaxy color-magnitude 
relation slope, but we can measure the scatter in color
of distant red field galaxies.  We find a scatter of 0.11 mag in restframe
$U-B$, even if only spheroidal types are included.  
This scatter is comparable to that found in COMBO-17 (B03),
and in a sample of red galaxies selected from the RC3, but 
larger than found in local or high-redshift early-type 
{\it cluster} galaxies
(Bower \etal\ 1992; van Dokkum \etal\ 2000).

DEEP1 galaxies have structural measurements from HST imaging of the
Groth Strip (Simard \etal\ 2002).  Red galaxies are generally 
more centrally concentrated than blue galaxies, as found
at low redshift in the SDSS (Blanton \etal\ 2003).  The bulk of
red galaxies have red stellar bulges, while blue galaxies 
generally have exponential profiles.  Red bulges are brighter
than blue starforming centers, so that the same problem of
progenitors occurs for red bulges as for red galaxies in toto.

We have classified red galaxies into three main groups: spheroidal red
galaxies (SRGs), edge-on disks, and diffuse red galaxies (DIFRGs).
These classifications correlate well with objective measures of
central concentration.  The SRGs include objects which resemble local
spheroid-dominated galaxies of type E, S0 and Sa; these are presumably
red due to old, metal-rich stellar populations.  The edge-on disks are
presumably reddened by dust.  

The DIFRGs are the most puzzling category: these do not
have a strong central concentration, are not edge-on,
and resemble local diffuse
galaxies such as Magellanic irregulars, but of course are both more
luminous and redder.  They are not obviously interacting or merging.
Our measurements of line emission from the DIFRGs indicate that they are
star-forming and likely reddened by dust rather than old stellar
populations.  Based on an
examination of red galaxies in a local sample drawn from the RC3,
there is no common type of local red galaxy with morphology 
similar to the DIFRGs; M82 may be the best analog.  The GEMS survey of
red galaxy morphologies at $0.65<z<0.75$ did not note DIFRGs as a class
(Bell \etal\ 2003b), but this may be influenced by the relatively 
low redshift range studied.
The DIFRGs in the DEEP1 sample are all at $z>0.65$, above the median
redshift of our survey, suggesting that DIFRGs are more common at 
high redshift.  Studies of extremely red objects (EROs) in the 
GOODS-S field indeed find fewer spheroidals and a larger fraction of 
irregular and ``other'' types at $z>1.5$ (Moustakas \etal\ 2004).
If in fact DIFRGs are similar to dusty EROs, they provide an 
opportunity to study the ERO phenomenon at lower redshifts, where
nebular features are accessible to optical spectroscopy.
%The question of what dusty EROs and DIFRGs evolve into remains
%to be explored.

This paper has focused mainly on the properties of distant
red-population galaxies.  Principal unanswered questions concerning 
them are the color evolution of red galaxies,
the manner in which galaxies evolve onto the red sequence,
how their numbers and total stellar mass change with time,
and the nature of diffuse red galaxies and their descendants.
Though blue galaxies are not highlighted here, the evolution in
their properties may be even greater than that of red galaxies,
given their color evolution.  A worthy project for the future will
be correlating the color evolution of blue galaxies
with other properties such as 
mass, star formation rate, and structure, and studying the entire
suite of properties as a function of location in the color-magnitude
diagram.

\medskip
\quad

\acknowledgments
The authors thank the staffs of HST and Keck for their
help in acquiring the spectroscopic data, the W. M. 
Keck Foundation and NASA for construction of the Keck telescopes,
and Bev Oke and Judy Cohen for their tireless work on LRIS that 
enabled the spectroscopic observations.  
The authors wish to recognize and acknowledge the 
cultural role and reverence that the summit of Mauna Kea
has always had within the indigenous Hawaiian community.
We are most fortunate to have the opportunity to conduct
observations from this mountain.  The DEEP surveys were founded
under the auspices of the NSF Center for Particle
Astrophysics.  The bulk of the work was supported by   
National Science Foundation grants AST 95-29098 and
00-71198 to UCSC.  Additional support came from
NASA grants AR-05801.01, AR-06402.01, and
AR-07532.01 from the Space Telescope Science Institute,
which is operated by AURA, Inc., under NASA contract NAS
5-26555.  HST imaging of the Groth Strip was planned,
executed, and analyzed by Ed Groth and Jason Rhodes with
support from NASA grants NAS5-1661 and NAG5-6279 from the WFPC1 IDT. 
SMF would like to thank the California
Association for Research in Astronomy for a generous
research grant.  NPV acknowledges support from
NASA grant GO-07883.01-96A, and NSF grants
NSF-0349155 from the Career Awards Program and NSF-0123690 via
the ADVANCE Institutional Transformation Program at NMSU.
We thank Nick Kaiser, Gerry Luppino 
and Alison Coil for the use of their photometric catalog,
Eric Bell and Chris Wolf for valuable discussions and 
information on the COMBO-17 
survey, and Jason X. Prochaska for a reading of the manuscript.

% \label{sec-references}

\end{document}